 	\def\version{September 4, 2005}
\newif\ifpdf
\ifx\pdfoutput\undefined
\pdffalse % we are not running PDFLaTeX
\else
\pdfoutput=1 % we are running PDFLaTeX
\pdftrue \fi

\newif\iffinal
\finalfalse	% Not final version
\finaltrue % Final version

\documentclass[reqno,twoside,11pt]{amsart}
\iffinal\else\usepackage[notref,notcite]{showkeys}\fi
\usepackage{amsmath}
\usepackage{amsfonts}
\usepackage{amssymb}
\usepackage{graphicx}
\usepackage{graphics}
\usepackage{verbatim}

\IfFileExists{epsf.def}{\input epsf.def}{\usepackage{epsf}}

\IfFileExists{myowntimes.sty}
	{\usepackage{myowntimes}\usepackage{mathrsfs}}
	{\usepackage{times}\IfFileExists{mathrsfs.sty}
		{\usepackage{mathrsfs}}{\newcommand{\mathscr}{\mathcal}}
	}
\DeclareFontFamily{OT1}{eusb}{} \DeclareFontShape{OT1}{eusb}{m}{n} {<5> <6> <7> <8> <9> <10> <11> <12> <14.4> eusb10}{}
\DeclareMathAlphabet{\eusb}{OT1}{eusb}{m}{n}

\DeclareFontFamily{OT1}{eusm}{} \DeclareFontShape{OT1}{eusm}{m}{n} {<5> <6> <7> <8> <9> <10> <11> <12> <14.4> eusm10}{}
\DeclareMathAlphabet{\eusm}{OT1}{eusm}{m}{n}

\DeclareFontFamily{OT1}{eufm}{} \DeclareFontShape{OT1}{eufm}{m}{n} {<5> <6> <7> <8> <9> <10> <11> <12> <14.4> eufm10}{}
\DeclareMathAlphabet{\mathfrak}{OT1}{eufm}{m}{n}

\DeclareFontFamily{OT1}{fraktura}{}
\DeclareFontShape{OT1}{fraktura}{m}{n} {<5> <6> <7> <8> <9> <10> <11> <12> <13> <14.4> [1.1] eufm10}{}
\DeclareMathAlphabet{\fraktura}{OT1}{fraktura}{m}{n}

\DeclareFontFamily{OT1}{cmfi}{} \DeclareFontShape{OT1}{cmfi}{m}{n} {<5> <6> <7> <8> <9> <10> <11> <12> <13> <14.4> [0.9] cmfi10}{}
\DeclareMathAlphabet{\cmfi}{OT1}{cmfi}{b}{n}

\DeclareFontFamily{OT1}{cmss}{} \DeclareFontShape{OT1}{cmss}{m}{n} {<5> <6> <7> <8> <9> <10> <11> <12> <13> <14.4> cmss10}{}
\DeclareMathAlphabet{\cmss}{OT1}{cmss}{m}{n}
\newtheoremstyle{thm}{1.5ex}{1.5ex}{\itshape\rmfamily}{} {\bfseries\rmfamily}{}{2ex}{}

\newtheoremstyle{def}{1.5ex}{1.5ex}{\rmfamily}{} {\bfseries\rmfamily}{}{2ex}{}

\newtheoremstyle{rem}{1.3ex}{1.3ex}{\rmfamily}{} {\itshape}
{} {1.5ex}{}

\newenvironment{proofsect}[1] {\vskip0.1cm\noindent{\rmfamily\itshape#1.}}{\qed\vspace{0.15cm}}%{\newline\vspace{0.15cm}}

\theoremstyle{thm}
\newtheorem{theorem}{Theorem}[section]
\newtheorem{lemma}[theorem]{Lemma}
\newtheorem{proposition}[theorem]{Proposition}

\newtheorem*{Main Theorem}{Main Theorem.}
\newtheorem{corollary}[theorem]{Corollary}

\theoremstyle{def}
\newtheorem{definition}[theorem]{Definition}

\theoremstyle{rem}
\newtheorem{remark}[theorem]{{\itshape Remark}}

\numberwithin{equation}{section}

\renewcommand{\section}{\secdef\sct\sect}
\newcommand{\sct}[2][default]{\refstepcounter{section}
\addcontentsline{toc}{section}
{{\tocsection {}{\thesection}{\!\!\!\!#1\dotfill}}{}}
\vspace{0.7cm}
\centerline{ %\large
\scshape\arabic{section}.\ #1} \nopagebreak \vspace{0.2cm}}
\newcommand{\sect}[1]{
\vspace{0.4cm} \centerline{\large\scshape\rmfamily #1}
\vspace{0.2cm}}

\renewcommand{\subsection}{\secdef\subsct\sbsect}
\newcommand{\subsct}[2][default]{\refstepcounter{subsection}
\addcontentsline{toc}{subsection}
{{\tocsection{\!\!}{\hspace{1.2em}\thesubsection}{\!\!\!\!#1\dotfill}}{}}
\nopagebreak\vspace{0.45\baselineskip} {\flushleft\bf
\arabic{section}.\arabic{subsection}~\bf #1.~}
\\*[3mm]\noindent
\nopagebreak}
\newcommand{\sbsect}[1]{\vspace{0.1cm}\noindent
\textbf{#1.~}\vspace{0.1cm}}

\renewcommand{\subsubsection}{%
\secdef \subsubsect\sbsbsect}
\newcommand{\subsubsect}[2][default]{%
\refstepcounter{subsubsection} 
\addcontentsline{toc}{subsubsection}{{\tocsection{\!\!}
{\hspace{3.05em}\thesubsubsection}{\!\!\!\!#1\dotfill}}{}}
\nopagebreak
\vspace{0.15\baselineskip} \nopagebreak {\flushleft\rmfamily
\itshape\arabic{section}.\arabic{subsection}.\arabic{subsubsection}
\ \rmfamily #1\/.}\ }
\newcommand{\sbsbsect}[1]{\vspace{0.1cm}\noindent
\rmfamily \itshape
\arabic{section}.\arabic{subsection}.\arabic{subsubsection} \
\sffamily #1\/.\ }

\iffinal
\newcommand{\printversion}{}
\else
\newcommand{\printversion}{, 
\version}
\fi

\newcommand{\diam}{\operatorname{diam}}

\newcommand{\textd}{\text{\rm d}\mkern0.5mu}

\newcommand{\texte}{\text{\rm e}\mkern0.7mu}

\renewcommand{\AA}{\mathcal A}
\newcommand{\BB}{\mathcal B}
\newcommand{\CC}{\mathcal C}
\newcommand{\DD}{\mathcal D}
\newcommand{\EE}{\mathcal E}

\newcommand{\GG}{\mathcal G}

\newcommand{\A}{\mathbb A}

\newcommand{\E}{\mathbb E}

\newcommand{\BbbP}{\mathbb P}

\newcommand{\R}{\mathbb R}

\newcommand{\T}{\mathbb T}

\newcommand{\Z}{\mathbb Z}

\newcommand{\scrC}{\mathscr{C}}

\newcommand{\scrF}{\mathscr{F}}

\newcommand{\twoeqref}[2]{(\ref{#1}--\ref{#2})}
\newcommand{\1}{{1\mkern-4.5mu\textrm{l}}}
\renewcommand{\1}{\text{\sf 1}}
\newcommand{\0}{\text{\sf 0}}

\newcommand{\bt}{\boldsymbol t}

\def\myffrac#1#2 in #3{\raise 2.6pt\hbox{$#3 #1$}\mkern-1.5mu\raise 0.8pt\hbox{$#3/$}\mkern-1.1mu\lower 1.5pt\hbox{$#3 #2$}}
\newcommand{\ffrac}[2]{\mathchoice%
	{\myffrac{#1}{#2} in \scriptstyle}
	{\myffrac{#1}{#2} in \scriptstyle}
	{\myffrac{#1}{#2} in \scriptscriptstyle}
	{\myffrac{#1}{#2} in \scriptscriptstyle}
}
\newcommand{\Vvert}{|\mkern-2.2mu|\mkern-2.2mu|}
\newcommand{\cc}{{\text{\rm c}}}
\newcommand{\pp}{\fraktura p}
\newcommand{\frakG}{\mathfrak G}
\newcommand{\frakS}{\mathfrak S}
\newcommand{\betat}{\beta_{\text{\rm t}}}
\newcommand{\Tt}{T_{\text{\rm t}}}

\newcommand{\bS}{\boldsymbol S}
\newcommand{\hate}{\hat{\text{\rm e}}}

\begin{document}
%\begin{comment}
\title[Forbidden gap argument and chessboard estimates\printversion] 
{\Large Forbidden gap argument for phase transitions\\proved by means of chessboard estimates}
%{\Large Weak completeness of phase diagram for phase transitions by chessboard estimates}

\author[M.~Biskup and R.~Koteck\'y\printversion] {Marek~Biskup$^1$\, and\, Roman Koteck\'y$^2$}

\thanks{\hglue-4.5mm\fontsize{9.6}{9.6}\selectfont\copyright\,2005 by M.~Biskup and R.~Koteck\'y. 
Reproduction, by any means, of the entire article for non-commercial purposes is permitted without charge.\vspace{2mm}}
\maketitle

\vspace{-5mm}
\centerline{\textit{$^1$Department of Mathematics, UCLA, Los Angeles, California, USA}}
\centerline{\textit{$^2$Center for Theoretical Study, Charles University, Prague, Czech Republic}}

\vspace{2mm}
\begin{quote}
\footnotesize \textbf{Abstract:} 
Chessboard estimates are one of the standard tools for proving phase coexistence in spin systems of physical interest.
In this note we show that the method not only produces a point in the phase diagram where more than one Gibbs states coexist, 
but that it can also be used to rule out the existence of shift-ergodic states that differ significantly from those proved to exist. 
 For models depending on a parameter (say, the temperature), this shows that the values of 
the conjugate thermodynamic quantity (the energy) inside the ``transitional gap'' are forbidden in all shift-ergodic Gibbs states. 
We point out several models where our result provides useful 
additional information concerning the set of possible thermodynamic equilibria.

\end{quote}
\vspace{2mm}

\section{Introduction}
\noindent
One of the basic tasks of mathematical statistical mechanics is to find a rigorous approach 
to various first-order phase transitions in lattice spin systems. Here two methods of proof are generally available: 
Pirogov-Sinai theory and chessboard estimates. The former, developed in \cite{PSa,PSb}, 
possesses an indisputable advantage of robustness with respect to (general) perturbations, 
but its drawbacks are the restrictions---not entirely without hope 
of being eventually eliminated~\cite{Imbrie1,Imbrie2,Dobrushin-Zahradnik,Zahradnik,BW1}---to (effectively) finite sets of possible spin values 
and to situations with rapidly decaying correlations. The latter method, which goes back to \cite{FL,FILS1,FILS2}, is limited, for the most part,  
to systems with nearest-neighbor interactions
but it poses almost no limitations on the individual spin space and/or the rate of correlation decay; see e.g.~\cite{Bruno-Messager}.

While both techniques ultimately produce a proof of phase coexistence, Pirogov-Sinai theory offers significantly better control 
of the number of possible Gibbs states.
Indeed, one can prove the so called \emph{completeness of phase diagram}~\cite{Z,BW2} which 
asserts that the states constructed by the theory exhaust the set of all shift-ergodic Gibbs states.
(In technical terms, there is
a one-to-one correspondence between the   shift-ergodic Gibbs states and the
``stable phases''  defined in terms of minimal ``metastable free energy''.)
Unfortunately, no conclusion of this kind is currently available in the approaches based solely on chessboard estimates.
This makes many of the conclusions of this technique---see~\cite{CSZ,SZ,BCKiv,BCN1,ES} for a modest 
sample of recent references---seem to be somewhat ``incomplete.''

To make the distinction more explicit, let us consider the example of temperature-driven first-order phase transition
in the~$q$-state Potts model with~$q\gg1$.
In dimensions~$d\ge2$, there exists a transition temperature,~$\Tt$, 
at which there are~$q$ ordered states that are low on both entropy and energy, 
and one disordered state which is abundant in both quantities.
The transition is accompanied by a massive jump in the energy density (as a function of temperature).
Here the ``standard'' proof based on chessboard estimates~\cite{Kotecky-Shlosman,KS-proceedings} produces ``only'' the existence of a temperature 
where the aforementioned~$q+1$ states coexist, but it does not rule out the existence of other states; particularly, 
those with energies ``inside'' the jump.
On the other hand, Pirogov-Sinai approaches~\cite{KLMR,LMMRS} permit us to conclude 
that \emph{no other} than the above~$q+1$ shift-ergodic Gibbs states can exist at~$\Tt$ and, in particular, 
there is a \emph{forbidden gap} of energy densities where no shift ergodic Gibbs states are allowed to enter.

The purpose of this note is to show that, after all, chessboard estimates can also be supplemented with 
a corresponding ``forbidden-gap'' argument. 
Explicitly, we will show that the calculations (and the assumptions) used, e.g., 
in~\cite{Kotecky-Shlosman,CSZ, SZ, BCKiv, BCN1,ES} to prove the \emph{existence} 
of particular Gibbs states at the corresponding transition temperature, or other driving parameter, 
imply also the \emph{absence} of Gibbs states that differ significantly from those proved to exist. 
We emphasize that no statement about the \emph{number} of possible extremal, translation-invariant Gibbs states is being made here, 
i.e., the  completeness of phase diagram in its full extent remains unproved. Notwithstanding, our results go some way towards a proof 
of completeness by ruling out, on general grounds, all but a ``small neighborhood'' of the few desired states (which may themselves be a non-trivial convex combination of extremal states).

The assumptions we make are quite modest; indeed, apart from the necessary condition of reflection positivity we require only 
translation invariance 
and absolute summability of interactions.  
And, of course, the validity---uniformly in the parameter driving the transition---of a bound that is generally used to suppress the contours 
while proving  the existence of coexisting phases. 
We also remark that the conclusion 
about the ``forbidden gap'' should not be interpreted too literally. Indeed, there are systems 
(e.g., the Potts model in an external field) where more than one gap  may ``open up'' at the transition. 
Obviously, in such situations one may have to consider a larger set of observables and/or richer parametrization of the model. 
We refer the reader to our theorems for the precise interpretation of the phrase ``forbidden gap'' in a general context.

The main idea of the proof is that all Gibbs states (at the same temperature) have the same large-deviation properties on the scale 
that is exponential in volume. This permits us to compare any translation-invariant Gibbs state with a corresponding measure on the torus, 
where chessboard estimates can be used to rule out most of the undesirable scenarios.
The comparison with torus boundary conditions requires a estimate on the interaction ``across'' the boundary; 
as usual this is implied by the absolute summability of interactions.
This is the setting we assume for the bulk of this paper (cf Theorem~\ref{T:complete}). 
For systems with unbounded interactions, a similar conclusion can be made under the assumption that the interactions are integrable 
with respect to the measures of interest (see Theorem~\ref{thm4.1}).

The rest of this paper is organized as follows:
In Sect.~\ref{sec2.1} and~\ref{sec2.2} we define the class of models to which our techniques apply and review various elementary facts 
about reflection positivity and chessboard estimates.
The statements of our main theorems (Theorem~\ref{T:complete} and Corollary~\ref{C:coex}) come in Sect~\ref{sec2.3}.
The proofs constitute the bulk of Sect.~\ref{sec3}; 
applications to recent results established by means of chessboard estimates are discussed in Sect.~\ref{sec4}.
The Appendix (Sect.~\ref{sec5}) contains the proof of Theorem~\ref{thm4.1} which provides an explicit estimate on the energy gap 
from Theorem~3 of~\cite{ES}.
This result is needed for one of our applications in Sect.~\ref{sec4}.

\section{Main result}
\noindent
In order to formulate our principal claims we will first recall the standard setup for proofs of first-order phase transitions 
by chessboard estimates and introduce the necessary notations.
The actual theorems are stated in Sect.~\ref{sec2.3}.

\subsection{Models of interest}
\label{sec2.1}\noindent
We will work with the standard class of spin systems on~$\Z^d$ and so we will keep our discussion of general concepts at the minimum possible. 
We refer the reader to Georgii's monograph~\cite{Georgii} for a more comprehensive treatment and relevant references.

Our spins,~$s_x$, will take values in a compact separable metric space~$\Omega_0$.
We equip~$\Omega_0$ with the~$\sigma$-algebra~$\scrF_0$ of its Borel subsets and  consider an \emph{a priori} probability 
measure~$\nu_0$ on~$(\Omega_0,\scrF_0)$. 
Spin configurations on~$\Z^d$ are the collections~$(s_x)_{x\in \Z^d}$.
We will use~$\Omega=\Omega_0^{\Z^d}$ to denote the set of all spin configurations on~$\Z^d$ 
and~$\scrF$ to denote the~$\sigma$-algebra of Borel subsets of~$\Omega$ defined using the product topology.
If~$\Lambda\subset \Z^d$, we define~$\scrF_\Lambda$  to be the sub-$\sigma$-algebra of events
depending only on~$(s_x)_{x\in \Lambda}$. 
For each~$x\in\Z^d$, the map~$\tau_x\colon\Omega\to\Omega$ is the ``translation
by~$x$'' defined by~$(\tau_x s)_y=s_{x+y}$. 
It is easy to check that~$\tau_x$ is a continuous and hence measurable for all~$x\in\Z^d$.
We will write~$\Lambda\Subset\Z^d$ to indicate that~$\Lambda$ is a finite subset of~$\Z^d$. 

To define Gibbs measures, we will consider a family of Hamiltonians~$(H_\Lambda)_{\Lambda\Subset\Z^d}$. 
These will be defined in terms of interaction potentials~$(\Phi_A)_{A\Subset\Z^d}$. 
Namely, for each~$A\Subset\Z^d$, let~$\Phi_A\colon\Omega\to\mathbb R$ be a function with the following properties:
\settowidth{\leftmargini}{(111)}
\begin{enumerate}
\item[(1)]~The function $\Phi_A$ is~$\scrF_A$-measurable for each~$A\Subset\Z^d$.
\item[(2)]~The interaction $(\Phi_A)$  is translation invariant, i.e.,~$\Phi_{A+x}=\Phi_A\circ \tau_x$ 
for all~$x\in\Z^d$ and all~$A\Subset\Z^d$.
\item[(3)]~The interaction $(\Phi_A)$ is absolutely summable in the sense that
\begin{equation}
\label{Norm-Phi}
\Vvert\Phi\Vvert=\sum_{\begin{subarray}{c}
A\Subset\Z^d\\0\in A
\end{subarray}}
\Vert\Phi_{A}\Vert_\infty<\infty.
\end{equation}
\end{enumerate}
The Hamiltonian on a set~$\Lambda\Subset\Z^d$ is a function~$H_\Lambda\colon\Omega\to\R$ 
defined by
\begin{equation}
\label{E:Ham}
H_\Lambda = \sum_{\begin{subarray}{c}
A\Subset\Z^d\\
A\cap\Lambda \neq\emptyset
\end{subarray}}\Phi_{A}.
\end{equation}
For each~$\beta\ge 0$, let~$\frakG_\beta$ be the set of Gibbs measures for the Hamiltonian \eqref{E:Ham}.
Specifically, $\mu\in \frakG_\beta$ if and only if
the conditional probability~$\mu(\,\cdot\,|\scrF_{\Lambda^\cc})$---which exists 
since~$\Omega$ is a Polish space---satisfies, for all~$\Lambda\Subset\Z^d$ and~$\mu$-almost all~$s$, the (conditional) DLR equation
\begin{equation}
\label{E:DLR}
\mu(\textd s_\Lambda|\scrF_{\Lambda^\cc})(s)=\frac{\texte^{-\beta H_\Lambda(s)}}{Z_\Lambda}\prod_{x\in\Lambda}\nu_0(\textd s_x).
\end{equation}
Here~$Z_\Lambda=Z_\Lambda(\beta, s_{\Lambda^\cc})$ is a normalization constant which is independent  
of ~$s_\Lambda=(s_x)_{x\in\Lambda}$.

\begin{remark}
The results of the present paper can be generalized even to the situations with unbounded spins and interactions; see Theorem~\ref{thm4.2}. 
However, the general theory of Gibbs measures with unbounded spins features some unpleasant 
technicalities that would obscure the presentation.
We prefer to avoid them and to formulate  the bulk of the paper for systems with compact spins.
Our restriction to translation-invariant interactions in~(2) above is mostly for convenience of exposition.
Actually, the proofs in Sect.~\ref{sec3} can readily be modified to include periodic interactions as well.
\end{remark}

\subsection{Chessboard estimates}
\label{sec2.2}\noindent
As alluded to before, chessboard estimates are among the principal tools for proving phase coexistence.
In order to make this tool available, we have to place our spin system on a torus.
Let~$\mathbb T_L$ be the torus of~$L\times \dots\times L$ sites 
and let ~$H_L\colon\Omega_0^{\mathbb T_L}\to \mathbb R$ be the function defined as follows.
Given a configuration~$s=(s_x)_{x\in \mathbb T_L}$, we extend~$s$ periodically to a configuration~$\bar s$ on all of~$\Z^d$. 
Using~$H_{\mathbb T_L}$ to denote the Hamiltonian associated with the embedding of~$\mathbb T_L$
into~$\Z^d$, we define~$H_L(s)=H_{\mathbb T_L}(\bar s)$.
The \emph{torus measure}~${\mathbb P}_{L,\beta}$  then simply is
\begin{equation}
\label{E:muL}
{\mathbb P}_{L,\beta}(\textd s)=\frac{\texte^{-\beta H_L(s)}}{Z_L}\prod_{x\in\mathbb T_L}\nu_0(\textd s_x).
\end{equation}
Here~$Z_L=Z_L(\beta)$ is the torus partition function.

Chessboard estimates will be implied by the condition of \emph{reflection positivity}. 
While this condition can already be defined in terms of interactions~$(\Phi_\Lambda)_{\Lambda\Subset\Z^d}$,
it is often easier to check it directly on the torus.
Let us consider a torus~$\mathbb T_L$ with even~$L$ and let us split it 
into two symmetric halves,~$\mathbb T_L^+$ and~$\mathbb T_L^-$, sharing a ``plane of sites'' on their boundary.
We will refer to the set~$P=\mathbb T_L^+\cap\mathbb T_L^-$ as a \emph{plane of reflection}.
Let~$\scrF_P^+$ and ~$\scrF_P^-$  denote the~$\sigma$-algebras of events depending only on configurations 
in~$\mathbb T_L^+$ and~$\mathbb T_L^-$,  respectively.

We assume that the naturally-defined (spatial) reflection~$\vartheta_P\colon\mathbb T_L^+ \leftrightarrow\mathbb T_L^-$
gives rise to a map $\theta_P\colon\Omega_0^{\mathbb T_L}\to\Omega_0^{\mathbb T_L}$
which obeys the following constraints:
\begin{enumerate}
\item[(1)]~$\theta_P$ is an \emph{involution},~$\theta_P\circ\theta_P=\operatorname{id}$.
\item[(2)]~$\theta_P$ is a \emph{reflection} in the sense that if~$\AA\in\scrF_P^+$ depends 
only on configurations in ~$\Lambda\subset\mathbb T_L^+$, then~$\theta_P(\AA)\in \scrF_P^-$ 
depends only on configurations in~$\vartheta_P(\Lambda)$.
\end{enumerate}
In many cases of interest,~$\theta_P$ is simply the mapping  that is directly induced by  the spatial reflection~$\vartheta_P$, 
i.e.,~$\theta_P=\vartheta_P^*$, where~$\bigl(\vartheta_P^*(s)\bigr)_x=s_{\vartheta_P(x)}$; 
our definition  permits us to combine the spatial reflection with an involution
of the single-spin space.

\smallskip
Reflection positivity is now defined as follows:

\begin{definition}
\label{D:RP}
Let~$\BbbP$ be a probability measure on~$\Omega_0^{\mathbb T_L}$ and let~$\E$ be the corresponding expectation.
We say that~$\BbbP$ is \emph{reflection positive}, if for any plane of reflection~$P$ 
and any two bounded $\scrF_P^+$-measurable random variables~$X$ and~$Y$,
\begin{equation}
\label{E:XY}
\E\bigl(X\theta_P(Y)\bigr)=\E\bigl(Y\theta_P(X)\bigr)
\end{equation}
and
\begin{equation}
\label{E:X}
\E\bigl(X\theta_P(X)\bigr)\ge 0.
\end{equation}
Here,~$\theta_P(X)$ denotes the $\scrF_L^-$-measurable random variable~$X\circ \theta_P$.
\end{definition}

\begin{remark}
Here are some standard examples of summable two-body interactions that are reflection positive.
Consider spin systems with vector-valued spins~$s_x$ and interaction potentials
\begin{equation}
%\label{}
\Phi_{\{x,y\}}=J_{x,y}\,(s_x, s_y),\qquad x\ne y,
\end{equation}
where~$J_{x,y}$ are coupling constants and $(\cdot,\cdot)$ denotes 
a positive-semidefinite inner product on~$\Omega$.
Then the corresponding torus Gibbs measure with~$\beta\ge0$ is reflection positive (for reflections through sites) for the following choices of~$J_{x,y}$'s:
\settowidth{\leftmargini}{(1111a)}
\begin{enumerate}
\item[(1)]
\emph{``Cube'' interactions}: Reflection-symmetric~$J_{x,y}$'s such that $J_{x,y}=0$ unless $x$ and~$y$ are vertices of a cube of~$2\times\dots\times2$ sites in~$\Z^d$.
\item[(2)]
\emph{Yukawa-type potentials}:
\begin{equation}
\label{1.2a}
J_{x,y}=\texte^{-\mu\vert x-y\vert_1},
\end{equation}
where~$\mu>0$ and $\vert x-y\vert_1$ is the $\ell^1$-distance between~$x$ and~$y$.
\item[(3)]
\emph{Power-law decaying interactions}:
\begin{equation}
\label{1.3a}
J_{x,y}=\frac1{\vert x-y\vert_1^{\varkappa}},
\end{equation}
with~$\varkappa> 0$.

\end{enumerate}
The proofs of these are based on the general theory developed in~\cite{FL,FILS1,FILS2}; 
relevant calculations can  also be found in~\cite[Sect.~4.2]{BCC}.
Of course, any linear combination of the above---as well as other reflection-positive interactions---with positive coefficients 
is still reflection positive.
\end{remark}

Now, we are finally getting to the setup underlying chessboard estimates.
Suppose that~$L$ is an integer multiple of an (integer) number~$B$. 
(To rule out various technical complications with the following theorem, we will actually always assume that~$\ffrac LB$ is a power of~$2$.)
Let~$\Lambda_B\subset \mathbb T_L$ be the box of~$(B+1)\times\dots\times (B+1)$ sites 
with the ``lower-left'' corner at the origin---we will call such box a~\emph{$B$-block}.
We can tile~$\mathbb T_L$ by translates of~$\Lambda_B$ by~$B$-multiples 
of vectors from the \emph{factor torus},~$\widetilde\T=\T_{L/B}$.
Note that the neighboring translates of~$\Lambda_B$ will have a side in common.
Let~$\AA$ be an event depending only on configurations in~$\Lambda_B$; we will call such~$\AA$ a \emph{$B$-block event}.
For each~$\bt\in \widetilde\T$, we define the event~$\theta_{\bt}(\AA)$ as follows:
\begin{enumerate}
\item[(1)]
If~$\bt$ has all components even, then~$\theta_{\bt}(\AA)$  is simply the translation of~$\AA$ by vector~$B\bt$, 
i.e., $\theta_{\bt}(\AA)=\tau_{B\bt}^{-1}(\AA)=\{s\in\Omega_0^{\T_L}\colon\tau_{B\bt}(s)\in\AA\}$.
\item[(2)]
For the remaining ~$\bt\in \widetilde\T$, we first reflect~$\AA$ through the ``midplane'' 
of~$\Lambda_B$ in all directions whose component
of~$\bt$ is odd, and then translate the result by~$B\bt$ as before.
\end{enumerate}
Thus,~$\theta_{\bt}(\AA)$ will always depend only on configurations in the $B$-block~$\Lambda_B+B\bt$.

\smallskip
The desired consequence of reflection positivity is now stated as follows.

\begin{theorem}[Chessboard estimate]
\label{T:chess}
Let~$\BbbP$ be a measure on~$\Omega_0^{\mathbb T_L}$ which is reflection-positive 
with respect to~$\theta_P$. 
Then for any $B$-block events~$\AA_1,\dots,\AA_m$ and any distinct sites~$\bt_1,\dots,\bt_m\in\widetilde\T$, 
\begin{equation}
\label{E:chess}
\BbbP\Bigl(\,\bigcap_{j=1}^m\theta_{\bt_j}(\AA_j)\Bigr)\le
\prod_{j=1}^m \BbbP\Bigl(\,\bigcap_{\bt\in\widetilde\T}\theta_{\bt}(\AA_j)\Bigr)^{\ffrac1{|\widetilde\T|}}.
\end{equation}
\end{theorem}
\begin{proof}
See \cite[Theorem~2.2]{FL}.
\end{proof}

The moral of this result---whose proof is nothing more than an enhanced version 
of the Cauchy-Schwarz inequality applied to the inner product~$X,Y\mapsto \E(X\theta_P(Y))$---is that the probability 
of any number of events factorizes, as a bound, into the product of probabilities. 
This is particularly useful for contour estimates; of course, provided that the word contour refers to a collection of boxes 
on each of which some ``bad'' event occurs. 
Indeed, by \eqref{E:chess} the probability of a contour will automatically be suppressed exponentially in the number of constituting ``bad'' boxes.
 
\subsection{Main theorems}
\label{sec2.3}\noindent
For any $B$-block event~$\AA$,  we introduce the quantity
\begin{equation}
\label{E:eta}
\pp_\beta(\AA)=\lim_{L\to\infty} \biggl(\mathbb P_{L,\beta}
\Bigl(\,\bigcap_{\bt\in\widetilde\T}\theta_{\bt}(\AA)\Bigr)\biggr)^{\ffrac1{|\widetilde\T|}},
\end{equation}
with the limit taken over multiples of~$B$. The limit exists by standard subadditivity arguments.
While the definition would suggest that~$\pp_\beta(\AA)$ is a large-deviation rate, 
chessboard estimates \eqref{E:chess} show that $\pp_\beta(\AA)$ can also be thought of as 
the ``probability of~$\AA$ regardless of the status of all other $B$-blocks.'' 
This interpretation is supported by the fact that~$\AA\mapsto\pp_\beta(\AA)$ is 
an outer measure on~$\scrF_{\Lambda_B}$ with~$\pp_\beta(\Omega)=1$, cf.~Lemma~6.3 of~\cite{BCN1}.

Furthermore, recalling that~$\Lambda_{N-1}$ is the block of~$N\times\cdots\times N$ sites with the ``lower-left'' corner at the lattice origin,
let
\begin{equation}
\label{E:R}
R_N(\AA)=\frac1{|\Lambda_{N-1}|}\sum_{x\in\Lambda_{N-1}}\1_\AA\circ\tau_{Bx}
\end{equation}
be the fraction of $B$-blocks (in $\Lambda_{NB-1}$) in which~$\AA$ occurs. 
Whenever~$\mu\in\frakG_\beta$ is a  Gibbs state for the Hamiltonian~\eqref{E:Ham} 
at inverse temperature~$\beta$ that is invariant with respect to the shifts~$(\tau_{Bx})_{x\in\Z^d}$, the limit
\begin{equation}
\label{E:rho}
\rho_\mu(\AA)=\lim_{N\to\infty} R_N(\AA)
\end{equation}
 exists~$\mu$-almost surely. 
In the following, we will use~$\rho_\mu(\AA)$ mostly for measures that are actually ergodic with respect to the shifts by multiples of~$B$.
In such cases the limit is self-averaging, $\rho_\mu(\AA)=\mu(\AA)$ almost surely.
Notwithstanding, we will stick to the notation~$\rho_\mu(\AA)$ to indicate that claims are being made about 
almost-sure properties of configurations and not just expectations.
To keep our statements concise, we will refer to measures which are invariant and ergodic 
with respect to the translations~$(\tau_{Bx})_{x\in\Z^d}$ as \emph{$B$-shift~ergodic}.

%%% ADDED A VERBAL EXPLANATORY SENTENCE TO THE THEOREM
\begin{comment}
Introducing a finite number of suitably chosen ``good''  $B$-block events $\GG_1, \dots, \GG_r$,
we will investigate the possible values for  the vector 
$\rho_\mu(\boldsymbol{\GG})=(\rho_\mu(\GG_i), i=1,\dots,n)\in [0,1]^n$
for all $B$-shift ergodic Gibbs states~$\mu\in\frakG_\beta$.  
Under appropriate assumptions, we get a general form of the  forbidden gap claim stating that
the vector  $\rho_\mu(\boldsymbol{\GG})$ always falls into a small neighborhood of one of coordinate unit vectors
from 
\begin{equation}
\label{E:}
\boldsymbol{E}=\{e_1,e_2,\dots,e_n\}\equiv \{(1,0,\dots,0),(0,1,\dots,0),\dots,(0,0,\dots,1)\}.
\end{equation}
\end{comment}
%%%%%%%%%%%%%

\smallskip
Our principal result can be formulated  as follows:

\begin{theorem}
\label{T:complete}
Let~$d\ge2$ and consider a spin system as described above for which that the torus measure is reflection positive 
for all~$\beta\ge 0$ and all even~$L\ge 2$. Let~$\GG_1, \dots, \GG_r$ be  a finite number of   $B$-block events
and let~$\BB=(\GG_1\cup\dots \cup \GG_r)^{\cc}$.
Suppose that the good block events are mutually exclusive and non-compatible 
(different types of goodness cannot occur in neighboring blocks): 
\begin{enumerate}
\item[(1)]~$\GG_i\cap\GG_j = \emptyset\ $ for all~$i\neq j$.
\item[(2)]
If~$\bt_1,\bt_2\in\widetilde\T$ are nearest neighbors, then
\begin{equation}
\label{E:non-comp}
\theta_{\bt_1}(\GG_i)\cap\theta_{\bt_2}(\GG_j )=\emptyset\  \text{ for all } i\neq j.
\end{equation}
\end{enumerate}
Then for every~$\epsilon>0$, there exists~$\delta>0$---which may depend on~$d$ but not on the details of the model nor on~$B$ or~$n$---such that for any~$\beta\ge 0$ with~$\pp_\beta(\BB)<\delta$ we have
\begin{equation}
\label{E:alt1}
\rho_\mu(\BB)\in [0,\epsilon]
\end{equation}
and
\begin{equation}
\label{E:alt2}
\rho_\mu(\GG_i)\in [0,\epsilon]\,\cup\,[1-\epsilon,1],\qquad i=1,\dots,r,
\end{equation}
for every $B$-shift ergodic Gibbs state~$\mu\in\frakG_\beta$. In particular, if~$\epsilon<\ffrac12$ then for every such~$\mu$ there exists a unique~$i$ such that~$\rho_\mu(\GG_i)\ge1-\epsilon$ and~$\rho_\mu(\GG_j)\le\epsilon$ for all~$j\ne i$.
\end{theorem}

We remark that the conclusion of Theorem~\ref{T:complete} holds even when the requirement of compact single-spin space 
and norm-bounded interactions are relaxed to the condition of \emph{finite average energy}. 
We state the corresponding generalization in Theorem~\ref{thm4.2}.
Theorem~\ref{T:complete} directly implies the standard conclusion of chessboard estimates 
(cf.~\cite[Propositions~3.1-3.3]{DS} or \cite[Theorem~4]{Kotecky-Shlosman}):
 
\begin{corollary}
\label{C:coex}
Let~$d\ge2$, let~$\beta_1<\beta_2$ be two inverse temperatures and let~$\GG_1$ and~$\GG_2$ be two mutually exclusive, non-compatible good $B$-block events  (cf conditions (1) and~(2) in Theorem~\ref{T:complete}). 
Then, for every~$\epsilon>0$ there exists a constant~$\delta>0$---which may depend on~$d$ but not~$B$ or the details of the model---such that the conditions
%\settowidth{\leftmargini}{(11)}
\begin{enumerate}
\item[(1)]~$\pp_\beta(\BB)<\delta$ for all~$\beta\in[\beta_1,\beta_2]$ and
\item[(2)]~$\pp_{\beta_1}(\GG_2)<\delta$ and~$\pp_{\beta_2}(\GG_1)<\delta$
\end{enumerate}
imply an  existence of an inverse temperature~$\betat\in(\beta_1,\beta_2)$ and of two distinct 
$B$-shift ergodic Gibbs measures~$\mu_1,\mu_2\in\frakG_{\betat}$ such that
\begin{equation}
\rho_{\mu_j}(\GG_j)\ge1-\epsilon,\qquad j=1,2.
\end{equation}
\end{corollary}

The above assumptions~(1) and~(2) appear in some form in all existing proofs based on chessboard estimates;
see Sect.~\ref{sec4} for some explicit examples.
The conclusions about the set of coexistence points can be significantly strengthened when, 
on the basis of thermodynamic arguments and/or stochastic domination, the expected amount 
of goodness~$\GG_2$ increases (and~$\GG_1$ decreases) with increasing~$\beta$.
For~$\epsilon\ll1$ the phase diagram then features a unique (massive) jump at some~$\betat$ 
from states dominated by~$\GG_1$ to those dominated by~$\GG_2$. Theorem~\ref{T:complete} 
implies that the bulk of the values inside the jump are not found in any ergodic Gibbs state.
Both Theorem~\ref{T:complete} and Corollary~\ref{C:coex} are proved in Sect.~\ref{sec3.2}.

\begin{remark}
Both results above single out inverse temperature as the principal parameter of interest. However, this is only a matter of convenience; all results hold equally well for any parameter of the model. An inspection of the proof shows that we can take~$\delta=c(d)\epsilon^{2/d}$ in Theorem~\ref{T:complete}, where~$c(d)$ is a constant that grows with dimension. However, the dependence on~$\epsilon$ should be significantly better; we made no attempts to reach the optimum. In any case, the fact that~$\delta$ does not depend on the details of the model is definitely sufficient to prove phase coexistence.
\end{remark}

\section{Proofs of main results}
\label{sec3}\noindent
We will assume that there is an ergodic Gibbs measure~$\mu\in\frakG_\beta$ that violates one of the conditions 
\twoeqref{E:alt1}{E:alt2}, and derive a contradiction.
Various steps of the proof will be encapsulated in technical lemmas in Sect.~\ref{sec3.1}; the actual proofs come in Sect.~\ref{sec3.2}.

\subsection{Technical lemmas}
\label{sec3.1}\noindent
Our first step is to convert the information about infinite-volume densities into a finite volume event.
Using the sites from~$\Lambda_{N-1}$ to translate the $B$-block~$\Lambda_B$ by multiples of~$B$ in each coordinate direction,
we get $\bigcup_{x\in\Lambda_{N-1}}(\Lambda_B+Bx)=\Lambda_{NB}$.
Similarly,  considering translates of~$\Lambda_{NB}$ by vectors~$NBx$ where~$x\in\Lambda_{M-1}$,
we get  $\bigcup_{x\in\Lambda_{M-1}}(\Lambda_{NB}+NBx)=\Lambda_{MNB}$ .
The important point is that, while the neighboring translates~$\Lambda_{NB}+NBx$ and~$\Lambda_{NB}+NBy$ are not disjoint,
they have only one of their $(d-1)$-dimensional sides in common.

Let~$\BB_N$ and~$\EE_{j,N}$, $j=1,\dots,r$, be events defined by
\begin{equation}
%\label{}
\BB_N=\bigl\{R_{N}(\BB)>\epsilon\bigr\}
\end{equation}
and
\begin{equation}
%\label{}
\EE_{j,N}=\bigl\{R_{N}({\GG_j})>\epsilon\bigr\},
\qquad j=1,\dots,r.
\end{equation}
Introducing the event
\begin{equation}
%\label{}
\EE_N=\BB_N\cup\bigcup_{1\le i<j\le r}(\EE_{i,N}\cap\EE_{j,N})
\end{equation}
and the fraction $R_{M,N}(\EE_N)$ of $BN$-blocks (in $\Lambda_{MNB}$) in which $\EE_N$ occurs,
\begin{equation}
\label{E:R_MN}
R_{M,N}(\EE_N)=\frac1{|\Lambda_{M-1}|}\sum_{x\in\Lambda_{M-1}}\1_{\EE_N}\circ\tau_{NBx},
\end{equation}
we have:

\begin{lemma}
\label{lemma3.1}
Let~$\epsilon<\ffrac12$ and consider a $B$-shift ergodic Gibbs measure~$\mu\in\frakG_\beta$ that violates one of the conditions 
\twoeqref{E:alt1}{E:alt2}. Then there exists an~$N_0<\infty$ and, for each~$N\ge N_0$, 
there exists an~$M_0=M_0(N)$ such that for all~$N\ge N_0$ and all~$M\ge M_0(N)$, one has
\begin{equation}
\label{E:R_MN>}
\mu\bigl(\,R_{M,N}(\EE_N)>\ffrac12\bigr)>\frac1{2N^d}.
\end{equation}
\end{lemma}

\begin{proofsect}{Proof}
The proof is based on a two-fold application of the Pointwise Ergodic Theorem.
Indeed, by ergodicity of~$\mu$ and Fatou's lemma we know that
\begin{equation}
%\label{}
\liminf_{N\to\infty}\mu(\BB_N)\ge\mu\bigl(\,\rho_\mu(\BB)>\epsilon\bigr)
\end{equation}
and
\begin{equation}
%\label{}
\liminf_{N\to\infty}\mu(\EE_{i,N}\cap\EE_{j,N})\ge\mu\bigl(\,\bigl\{\rho_\mu(\GG_i)>\epsilon\bigr\}\cap
\bigl\{\rho_\mu(\GG_j)>\epsilon\bigr\}\bigr).
\end{equation}
But~$\mu$ violates one of the conditions \twoeqref{E:alt1}{E:alt2} and so  either~$\rho_\mu(\BB)>\epsilon$ or
$\rho_\mu(\GG_i)>\epsilon$ \emph{and}~$\rho_\mu(\GG_j)>\epsilon$  for some~$i\ne j$.
All of these inequalities are valid~$\mu$-almost surely and so it follows that
\begin{equation}
%\label{}
\mu(\EE_N)\,\underset{N\to\infty}\longrightarrow\,1.
\end{equation}
Now, let us fix~$N$ so that~$\mu(\EE_N)\ge\ffrac34$. 
Then ergodicity with respect to translates by multiples of~$B$  
implies that
\begin{equation}
\begin{aligned}
\mu\Bigl(\,\bigcup_{y\in\Lambda_{N-1}}\bigl\{R_{M,N}(\EE_N)\circ\tau_{By}>\ffrac12\bigr\}\Bigr)
&\ge
\mu\Bigl(\,\frac1{N^d}\sum_{y\in\Lambda_{N-1}}R_{M,N}(\EE_N)\circ\tau_{By}>\frac12\Bigr)
\\
&=\mu\bigl(R_{MN}(\EE_N)>\ffrac12\bigr)
\,\underset{M\to\infty}\longrightarrow\,1.
\end{aligned}
\end{equation}
It follows that the left-hand side exceeds~$\ffrac12$ once~$M$ is sufficiently large, 
which in conjunction with subadditivity and~$\tau_{By}$-invariance of~$\mu$ directly implies~\eqref{E:R_MN>}.
\end{proofsect}

Our next task will be to express~$\EE_N$ solely in terms of conditions on bad $B$-blocks in~$\Lambda_{NB}=
\bigcup_{x\in\Lambda_{N-1}}(\Lambda_B+Bx)$.
Given two distinct sites~$x,y\in\Lambda_{N-1}$, let~$\{x\nleftrightarrow y\}$ 
denote the event that there is no nearest-neighbor path~$\pi=(x_1,\dots,x_k)$ on~$\Lambda_{N-1}$ such that
%\settowidth{\leftmargini}{(11)}
\begin{enumerate}
\item[(1)]
$\pi$ connects~$x$ to~$y$, i.e.,~$x_1=x$ and~$x_k=y$.
\item[(2)]
all~$B$-blocks ``along''~$\pi$ are good, i.e.,~$\tau_{Bx_j}(\BB^\cc)$ occurs for all~$j=1,\dots,k$.
\end{enumerate}
Note that~$\{x\nleftrightarrow y\}$ automatically holds when one of the blocks~$\Lambda_B+Bx$ or~$\Lambda_B+By$ is bad.
Further,  let $Y_N$ be the ($\scrF_{\Lambda_{NB}}$-measurable) random variable 
\begin{equation}
%\label{}
Y_N=\#\bigl\{(x,y)\in\Lambda_{N-1}\times\Lambda_{N-1}\colon x\ne y\,\&\,x\nleftrightarrow y\bigr\}
\end{equation}
and let~$\CC_N$  be the event
\begin{equation}
%\label{}
\CC_N=\bigl\{Y_N \ge (\epsilon N^d)^2\bigr\}.
\end{equation}
Conditions~(1) and~(2) from Theorem~\ref{T:complete} now directly imply:

\begin{lemma}
\label{L:incl}
For all~$N$, we have $\EE_N\subset\CC_N$.

\end{lemma}

\begin{proofsect}{Proof}
Clearly, we have $\BB_N\subset\CC_N$, and so we only have to show that
\begin{equation}
\label{E:EIEJ<CN}
\EE_{i,N}\cap\EE_{j,N}\subset\CC_N,\qquad 1\le i<j\le r.
\end{equation}
Let us fix~$i\ne j$ and recall that on~$\EE_{i,N}\cap\EE_{j,N}$, at least an~$\epsilon$-fraction of all $B$-blocks in~$\Lambda_{NB}$ 
will be~$i$-good and at least an~$\epsilon$-fraction of them will be~$j$-good.
By conditions~(1) and~(2) from Theorem~\ref{T:complete}, no two~$B$-blocks of different type of goodness can be connected 
by a path of good~$B$-blocks, and so there are at least~$(\epsilon N^d)^2$ pairs of distinct $B$-blocks in~$\Lambda_{NB}$ 
that are not connected to each other by a path of good blocks. This is exactly what defines the event~$\CC_N$. 
\end{proofsect}

The events~$\EE_N$ and~$\CC_N$ have the natural interpretation as~$NB$-block events on~$\T_L$ whenever~$L$ is divisible by~$NB$.
If~$\AA$ is such an~$NB$-block event, let~$\tilde\pp_\beta(\AA)$ denote the analogue of the quantity from \eqref{E:eta} 
where the $\theta_{\bt}$'s now involve translations by multiples of~$NB$.
Our next technical lemma provides an estimate on~$\tilde\pp_\beta(\CC_N)$ in terms of~$\pp_\beta(\BB)$:

\begin{lemma}
\label{lemma3.3}
Let~$d$ be the dimension of the underlying lattice and suppose that~$d\ge2$.
For each~$\epsilon>0$---underlying the definitions of~$\BB_N$, $\EE_N$ and~$\CC_N$---and each~$\eta>0$, 
there exists a number $\delta=\delta(\epsilon,\eta,d)>0$ such that if~$\pp_\beta(\BB)<\delta$, then $\tilde\pp_\beta(\CC_N)<\eta$.
\end{lemma}

\begin{proofsect}{Proof}
Let us use~$\Pi_{L,\beta}(\CC_N)$ to abbreviate the quantity
\begin{equation}
%\label{}
\Pi_{L,\beta}(\CC_N)=\BbbP_{L,\beta}\Bigl(\,\bigcap_{\bt\in\widetilde\T}\theta_{\bt}(\CC_N)\Bigr),
\end{equation}
where~$\widetilde\T=\T_{L/(NB)}$ is the factor torus in the present context.
Observing that~$\CC_N$ is preserved by reflections through the ``midplanes'' of~$\Lambda_{NB}$, 
a multivariate version of Chebyshev's inequality then yields
\begin{equation}
\label{3.15d}
\Pi_{L,\beta}(\CC_N)
\le\E_{L,\beta}\biggl(\,\prod_{\bt\in\widetilde\T}\frac{Y_N\circ\tau_{BN\bt}}{(\epsilon N^d)^2}\biggr).
\end{equation}
Here~$\E_{L,\beta}$ is the expectation with respect to~$\BbbP_{L,\beta}$.

To estimate the right-hand side of \eqref{3.15d}, we will rewrite~$Y_N$ as a sum.
Let~$x, y\in\Lambda_{N-1}$ be distinct. A connected subset $\Gamma\subset\Lambda_{N-1}$
is said to \emph{separate~$x$ from~$y$} (in $\Lambda_{N-1}$) if each nearest-neighbor path~$\pi$ from~$x$ to~$y$ 
on $\Lambda_{N-1}$ intersects~$\Gamma$.
We use $\frakS(x,y)$ to denote the set of all such sets $\Gamma\subset\Lambda_{N-1}$. 
Notice that~$\{x\},\{y\}\in \frakS(x,y)$.
We claim that, whenever $(x,y)$ is a pair of points contributing to~$Y_N$, there exists $\Gamma\in\frakS(x,y)$
separating~$x$ from~$y$ such that every block~$\Lambda_B+Bz$ with~$z\in\Gamma$ is bad.
Indeed, if $\Lambda_B+Bx$ is a bad block we take $\Gamma=\{x\}$.
If~$\Lambda_B+Bx$ is a good block, then we define~$\scrC_x$ to be the maximal connected subset of~$\Lambda_{N-1}$ 
containing~$x$ such that~$\Lambda_B+Bz$ is a good block for all~$z\in\scrC_x$, and let~$\Gamma$ be its external boundary.
Using~$\1_\Gamma$ to denote the indicator of the event that every block~$\Lambda_B+Bz$ with~$z\in\Gamma$ is bad, we get
\begin{equation}
%\label{}
Y_N\le\sum_{x,y\in\Lambda_{N-1}}\sum_{\Gamma\in\frakS(x,y)}
\1_\Gamma.
\end{equation}
Let~$K=(\frac L{BN})^d$ be the volume of the factor torus and let~$\bt_1,\dots,\bt_K$ be an ordering of all sites of~$\widetilde\T$.
Then we have
\begin{equation}
\label{E:Pi_L}
\Pi_{L,\beta}(\CC_N)\le\frac1{(\epsilon N^d)^{2K}}
\sum_{\begin{subarray}{c}
(x_j,y_j)\\ j=1,\dots, K
\end{subarray}}
\,\sum_{\Gamma_1,\dots,\Gamma_K}
\E_{L,\beta}\biggl(\,\prod_{j=1}^K\1_{\Gamma_j}\circ\tau_{BN\bt_j}\biggr),
\end{equation}
where  the first sum runs over collections of  pairs~$(x_j,y_j)$,  $j=1,\dots,K$,
 of distinct sites in~$\Lambda_{N-1}$ and the second sum is over all collections
of separating surfaces $\Gamma_j\in\frakS(x_j,y_j)$, $j=1,\dots,K$.

To estimate the right-hand side of \eqref{E:Pi_L} we define~$\pp_{L,\beta}(\BB)$ 
to be the quantity on the right-hand side of \eqref{E:eta},  before taking the limit~$L\to\infty$, with~$\AA=\BB$.
Since each indicator~$\1_{\Gamma_j}\circ\tau_{BN\bt_j}$ enforces bad blocks~$\Lambda_B+B(z+N\bt_j)$ for~$z\in\Gamma_j$, 
and the set of blocks
$\Lambda_B+B(z+N\bt_j)$,  $z\in\Lambda_{N-1}$, is, for $\bt_i\neq \bt_j$, disjoint from the set
$\Lambda_B+B(z+N\bt_i)$, $z\in\Lambda_{N-1}$, we can use chessboard estimates (Theorem~\ref{T:chess}) to get
\begin{equation}
%\label{}
\E_{L,\beta}\biggl(\,\prod_{j=1}^K\1_{\Gamma_j}\circ\tau_{BN\bt_j}\biggr)
\le\bigl[\pp_{L,\beta}(\BB)\bigr]^{|\Gamma_1|+\cdots+|\Gamma_K|}.
\end{equation}
A standard contour-counting argument now shows that, for any distinct~$x,y\in\Lambda_{N-1}$,
\begin{equation}
%\label{}
\sum_{\Gamma\in\frakS(x,y)}\bigl[\pp_{L,\beta}(\BB)\bigr]^{|\Gamma|}\le c_1\pp_{L,\beta}(\BB)^d
\end{equation}
with some constant~$c_1=c_1(d)$, provided that $\pp_{L,\beta}(\BB)$ is sufficiently small.
The sum over collections of  pairs~$(x_j,y_j)$,  $j=1,\dots,K$, contains  at most~$(N^{2d})^K$ terms,
 allowing us to bound
\begin{equation}
%\label{}
\Pi_{L,\beta}(\CC_N)\le\biggl(\frac{c_1\pp_{L,\beta}(\BB)^d}{\epsilon^2}\biggr)^K.
\end{equation}
Since~$\Pi_{L,\beta}(\CC_N)^{\ffrac1K}\to\tilde\pp_\beta(\CC_N)$ as~$L\to\infty$, 
it follows that~$\tilde\pp_\beta(\CC_N)\le c_1\pp_\beta(\BB)^d\epsilon^{-2}$, 
which for~$\pp_\beta(\BB)$ small enough, can be made smaller than any~$\eta$ initially prescribed.
\end{proofsect}

Our final technical ingredient is an estimate on the Radon-Nikodym derivative of a Gibbs measure~$\mu\in\frakG_\beta$ 
and the torus measure at the same temperature:

\begin{lemma}
\label{lemma3.4}
Let~$\Lambda_L\subset\Z^d$ be an~$L$-block and let~$\T_{2L}$ be a torus of side~$2L$. 
Let us view~$\Lambda_L$ as embedded into~$\T_{2L}$ and let~$\BbbP_{2L,\beta}$ be the torus Gibbs measure on~$\T_{2L}$.
Then for any~$a>0$ there exists~$L_0$ such that
\begin{equation}
%\label{}
\texte^{-\beta a L^d}\BbbP_{2L,\beta}(\AA)
\le\mu(\AA)\le 
\texte^{\beta aL^d}\BbbP_{2L,\beta}(\AA).
\end{equation}
  for all~$L\ge L_0$,  any~$\mu\in\frakG_\beta$, and any~$\scrF_{\Lambda_L}$-measurable event~$\AA$.
\end{lemma}

\begin{proofsect}{Proof}
For finite-range interactions, this lemma is completely standard. However, since our setting includes also interactions with infinite range, 
we provide a complete proof. We will prove only the right-hand side of the above inequality; the other side is completely analogous.

First, from the DLR equation we know that there exists a configuration~$s=(s_x)_{x\in\Z^d}$, such that
\begin{equation}
%\label{}
\mu(\AA|\scrF_{\Lambda^\cc})(s)\ge\mu(\AA)
\end{equation}
with the left-hand side of the form \eqref{E:DLR}.
Let~$s'$ be a configuration on~$\T_{2L}$.
We will show that $\mu(\,\cdot\,|\scrF_{\Lambda_L^\cc})(s)$ 
and $\BbbP_{2L,\beta}(\,\cdot\,|\scrF_{\Lambda_L^\cc})(s')$ are absolutely continuous 
with respect to each other---as measures on~$\scrF_{\Lambda_L}$---and the Radon-Nikodym derivative is bounded above 
by $\texte^{\beta a L^d}$ regardless of the ``boundary conditions''~$s$ and~$s'$.

Suppose that~$s'_x=s_x$ for all~$x\in\Lambda_L$ 
and let~$\bar s'$ be its~$2L$-periodic extension to all of~$\Z^d$.
Then the Radon-Nikodym derivative of $\BbbP_{2L,\beta}(\,\cdot\,|\scrF_{\Lambda_L^\cc})(s')$
with respect to the product measure $\prod_{x\in\Lambda_L}\nu_0(\textd s_x)$ is 
$\texte^{-\beta H_{\Lambda_L}(\bar s')}/Z_{\Lambda_L}(\bar s'_{\Lambda_L^\cc})$ while that 
of~$\mu(\,\cdot\,|\scrF_{\Lambda_L^\cc})(s)$ is $\texte^{-\beta H_{\Lambda_L}(s)}/Z_{\Lambda_L}(s_{\Lambda_L^\cc})$.
It thus suffices to show, uniformly in~$(s_x)_{x\in\Lambda_L}$, that
\begin{equation}
%\label{}
\bigl|H_{\Lambda_L}(s)-H_{\Lambda_L}(\bar s')\bigr|\le \frac a2 L^d
\end{equation}
once~$L$ is sufficiently large.
To this end, we first note that
\begin{equation}
\label{E:H-H}
\bigl|H_{\Lambda_L}(s)-H_{\Lambda_L}(\bar s')\bigr|\le 2\sum_{\begin{subarray}{c}
A\colon A\cap\Lambda_L\ne\emptyset\\A\cap\Lambda_L^\cc\ne\emptyset
\end{subarray}}
\Vert\Phi_{A}\Vert_\infty.
\end{equation}
To estimate the right-hand side, we will decompose~$\Lambda_L$ into ``shells,'' $\Lambda_n\setminus\Lambda_{n-1}$, and use 
the fact that if~$A$ intersects~$\Lambda_n\setminus\Lambda_{n-1}$ as well as~$\Lambda_L^\cc$, 
then the diameter of~$A$ must be at least~$L-n$.
Using the translation invariance of the interactions, we thus get
\begin{equation}
%\label{}
\sum_{\begin{subarray}{c}
A\colon A\cap\Lambda_L\ne\emptyset\\A\cap\Lambda_L^\cc\ne\emptyset
\end{subarray}}
\Vert\Phi_{A}\Vert_\infty
\le\sum_{n=1}^L|\Lambda_n\setminus\Lambda_{n-1}|
\,\sum_{\begin{subarray}{c}
A\colon 0\in A \\ \diam(A)\ge L-n
\end{subarray}}
\Vert\Phi_{A}\Vert_\infty.
\end{equation}
But~$\Vvert\Phi\Vvert<\infty$ implies that the second sum tends to zero as~$L-n\to\infty$ 
and since $|\Lambda_n\setminus\Lambda_{n-1}|=o(L^d)$ while $\sum_{1\le n\le L}|\Lambda_n\setminus\Lambda_{n-1}|=L^d$, 
the result is thus~$o(L^d)$.
In particular, for~$L$ sufficiently large, the right-hand side of \eqref{E:H-H} will be less than~$\frac{a}2L^d$.
\end{proofsect}

\subsection{Proofs of Theorem~\ref{T:complete} and  Corollary~\ref{C:coex}}
\label{sec3.2}\noindent
Now we are ready to prove our main theorem:

\begin{proofsect}{Proof of Theorem~\ref{T:complete}}
Fix~$\epsilon<\ffrac12$ and let~$\mu\in\frakG_\beta$ be a $B$-shift ergodic Gibbs measure for which one of the conditions 
\twoeqref{E:alt1}{E:alt2} fails. Applying Lemma~\ref{lemma3.1} and the inclusion in Lemma~\ref{L:incl} we find that
\begin{equation}
%\label{}
\mu\bigl(R_{M,N}(\CC_N)>\ffrac12\bigr)>\frac1{2N^d}
\end{equation}
once~$N\ge N_0$ and~$M\ge M_0(N)$. Now, consider the torus~$\T_L$ of side~$L=2MNB$ and 
embed~$\Lambda_{MNB}=\bigcup_{x\in\Lambda_{M-1}}(\Lambda_{NB}+NBx)$ into~$\T_L$ in the ``usual'' way.
By Lemma~\ref{lemma3.4} we know that for any \emph{fixed}~$N\ge N_0$, 
there exists a sequence~$a_M$ of positive numbers with~$a_M\downarrow 0$ as~$M\to\infty$, such that we have
\begin{equation}
\label{E:R_MN(C)>}
\BbbP_{L,\beta}\bigl(R_{M,N}(\CC_N)>\ffrac12\bigr)>\frac1{2N^d}
\texte^{-\beta(NB)^d a_M M^d},\qquad M\to\infty.
\end{equation}
Our goal is to show that, once~$N$ is chosen sufficiently large, the left-hand side is exponentially small in~$M^d$, 
thus arriving at a contradiction.

By conditioning on which of the~$M^d/2$ translates of~$\Lambda_{BN}$ have~$\CC_N$ satisfied, 
and applying the chessboard estimates in blocks of side~$NB$, we get
\begin{equation}
%\label{}
\BbbP_{L,\beta}\bigl(R_{M,N}(\CC_N)>\ffrac12\bigr)
\le 2^{M^d}\,\tilde\pp_{2L,\beta}(\CC_N)^{{M^d}/2},
\end{equation}
where~$\tilde\pp_{2L,\beta}(\CC_N)$ is the finite-torus version of~$\tilde\pp_\beta(\CC_N)$.
Next we choose~$\eta<\ffrac14$ and let~$\delta>0$ and~$N\ge N_0$ be such that the bounds in Lemma~\ref{lemma3.3} apply.
Then for all sufficiently large~$M$ (and hence all large~$L$) we have~$\tilde\pp_{2L,\beta}(\CC_N)<\eta$ and so
\begin{equation}
\label{3.29}
\BbbP_{L,\beta}\bigl(R_{M,N}(\CC_N)>\ffrac12\bigr)
\le (4\eta)^{{M^d}/2}.
\end{equation}
But this is true for all~$M\gg1$ and so the bound \eqref{E:R_MN(C)>} must be false.
Hence, no such~$\mu\in\frakG_\beta$ could exist to begin with; i.e., \twoeqref{E:alt1}{E:alt2} 
must hold for all $B$-shift ergodic~$\mu\in\frakG_\beta$.
\end{proofsect}

To finish our proofs, we will also need to establish our claims concerning phase coexistence:

\begin{proofsect}{Proof of Corollary~\ref{C:coex}}
Suppose that~$\epsilon$ and~$\delta$ are such that Theorem~\ref{T:complete} applies.
By condition~(1), the conclusions \twoeqref{E:alt1}{E:alt2} of this theorem are thus available for all~$\beta\in[\beta_1,\beta_2]$.
This implies
\begin{equation}
\label{3.28e}
\rho_\mu(\GG_j)\in[0,\epsilon]\,\cup\,[1-\epsilon,1],\qquad j=1,2,
\end{equation}
for every $B$-shift ergodic $\mu\in\frakG_\beta$ at every~$\beta\in[\beta_1,\beta_2]$.
We claim that~$\rho_\mu(\GG_2)$ is small in every ergodic state~$\mu\in\frakG_{\beta_1}$.  
Indeed, by Lemma~6.3 of~\cite{BCN1} and condition~(2) of the corollary, we have
\begin{equation}
%\label{}
\pp_{\beta_1}(\BB\cup\GG_2)\le \pp_{\beta_1}(\BB)+\pp_{\beta_1}(\GG_j)<2\delta.
\end{equation}
Hence, if the~$\delta$ in Corollary~\ref{C:coex} was so small that Theorem~\ref{T:complete} applies 
for some~$\epsilon<\ffrac12$ even when~$\delta$ is replaced by~$2\delta$,
we can regard $\BB\cup\GG_2$ as a bad event at~$\beta=\beta_1$ and conclude that~$\rho_\mu(\GG_2)<\ffrac12$, 
and hence $\rho_\mu(\GG_2)\le\epsilon$, by \eqref{3.28e}, in every ergodic~$\mu\in\frakG_{\beta_1}$.
A similar argument proves that~$\rho_\mu(\GG_1)\le\epsilon$ in every ergodic~$\mu\in\frakG_{\beta_2}$.
Usual weak-limit arguments then yield the existence of at least one point $\betat\in(\beta_1,\beta_2)$ 
where both types of goodness coexist.
\end{proofsect}

\section{Applications}
\label{sec4}\noindent
The formulation of our main result is somewhat abstract. 
In the present section, we will pick several models in which phase coexistence has been proved using chessboard estimates 
and use them to demonstrate the consequences of our main theorem. 
Although we will try to stay rather brief, we will show that, generally, 
the hypothesis of our main result---i.e., the assumption on smallness of the parameter $\pp_\beta(\BB)$---is directly implied 
by the calculations already carried out in the corresponding papers.
The reader should consult the original articles for more motivation and further details concerning the particular models.

\subsection{Potts model}
The~$q$-state Potts model serves as a paradigm of order-disorder transitions.
The existence of the transition has been proved by chessboard estimates in
\cite{Kotecky-Shlosman}.
While the completeness of the phase diagram has, in the meantime, been
established with the help of Pirogov-Sinai theory~\cite{Martirosian}, we find it useful to illustrate our 
general claims on this rather straightforward example.
Later on we will pass to more complex systems where no form of completeness---and, more relevantly, no ``forbidden gap''---has
been proved.

The spins~$\sigma_x$ of the $q$-state Potts model take values in the set~$\{1,\dots,q\}$ with \emph{a priori} equal probabilities.
The formal Hamiltonian is
 \begin{equation}
\label{E:Potts-Ham}
H(\sigma)=-\sum_{\langle  x,y \rangle}\delta_{\sigma_x,\sigma_y},
\end{equation}
where~$\langle x,y\rangle$ runs over all (unordered) nearest-neighbor pairs in~$\Z^d$.
The states of minimal energy have all neighboring spins equal, and so we expect that low temperature states are dominated 
by nearly constant spin-configurations.
On the other hand, at high temperatures the spins should be nearly independent and, in particular, 
neighboring spins will typically be different from each other.
This leads us to consider the following good events on $1$-block~$\Lambda_1$:
\begin{equation}
\label{E:P-ord-dis}
\begin{aligned}
\GG^{\text{dis}}&=\bigl\{\sigma\colon \sigma_x\neq \sigma_y \text{ for all }
x,y\in\Lambda_1, |x-y|=1\bigr\},
\\
\GG^{\text{ord},m}&=\bigl\{\sigma\colon \sigma_x=m \text{ for all } x\in
\Lambda_1\bigr\},\qquad m=1,\dots,q.
\end{aligned}
\end{equation}
Using similar events, it was proved~\cite{Kotecky-Shlosman} that, for $d\ge 2$ 
and~$q$ sufficiently large, there exists an inverse temperature~$\betat$ and~$q+1$ ergodic
Gibbs states~$\mu^{\text{dis}}\in\frakG_{\betat}$ and~$\mu^{\text{ord},m}\in\frakG_{\betat}$, $m=1,\dots,q$,
such that the corresponding 1-block densities satisfy
\begin{equation}
\label{E:dis}
\rho_{\mu^{\text{dis}}}(\GG^{\text{dis}})\ge 1-\epsilon
\end{equation}
and
\begin{equation}
\label{E:ord}
\rho_{\mu^{\text{ord},m}}(\GG^{\text{ord},m})\ge 1-\epsilon,
\qquad m=1,\dots,q,
\end{equation}
where $\epsilon=\epsilon(q)$ tends to zero as $q\to\infty$.
In addition, monotonicity of the energy density as a function of~$\beta$ can be invoked to show 
that $\rho_\mu(\GG^{\text{dis}})$ is large in all translation-invariant~$\mu\in\frakG_\beta$ when~$\beta<\betat$, 
while it is small in all such states when~$\beta>\betat$.

The full completeness \cite{Martirosian} asserts that the above-mentioned $q+1$ states exhaust the set of all 
shift-ergodic Gibbs states in $\mathfrak
G_{\betat}$.
A weaker claim follows as a straightforward application of our Theorem~\ref{T:complete}:
\emph{For each   shift-ergodic Gibbs state
$\mu\in\frakG_{\betat}$ there is  
either
$\rho_\mu(\GG^{\text{dis}})\ge 1-\epsilon$
or 
$\rho_\mu(\GG^{\text{ord},m})\ge 1-\epsilon $ for some}  $m=1,\dots,q$.

\noindent
The main hypothesis of our theorem amounts to the smallness of the quantity~$\pp_\beta(\BB)$,
where
\begin{equation}
%\label{}
\BB=\Bigl(\GG^{\text{dis}}\cup\bigcup_{m=1}^q\GG^{\text{ord},m}\Bigr)^\cc,
\end{equation}
which in turn boils down to an estimate on the probability of the disseminated event~$\BB$ on
the right-hand side of \eqref{E:eta}.
The needed estimate coincides with the bound provided in~\cite{Kotecky-Shlosman} 
by evaluating directly (i.e., ``by hand'') the energy and the number of contributing configurations.
The result---which in~\cite{Kotecky-Shlosman} appears right before the last formula on p.~506 is used 
to produce~(4.4$'$)---reads
\begin{equation}
\label{E:Potts-dissem-b}
\pp_\beta(\BB)\le
\Bigl[\frac{q^{d-2^{-(d-1)}}}{(q-2d)^d}\Bigr]^{\frac1{2d}}.
\end{equation}
This implies the needed bound once $q\gg1$.

\begin{remark}
Analogous calculations establish the corresponding forbidden gap in more complicated variants of the Potts model; see e.g.~\cite{BCK}.
\end{remark}

\subsection{Intermediate phases in dilute spin systems}
The first instance where our results provide some new insight are dilute annealed ferromagnets exhibiting staggered order phases 
at intermediate temperatures.
These systems have been studied in the context of both discrete  \cite{CKS1} and continuous spins \cite{CKS2}.
The characteristic examples of these classes are the \emph{site-diluted Potts model} with the Hamiltonian
\begin{equation}
\label{E:Potts-dil-Ham}
H(n,\sigma) = -\sum_{\langle x,y\rangle}n_x n_y(\delta_{\sigma_x, \sigma_y}-1) -\lambda \sum_x n_x - 
\kappa \sum_{\langle x,y\rangle}  n_x n_y
\end{equation}
and the \emph{site-diluted $XY$-model} with the Hamiltonian
\begin{equation}
\label{E:XY-dil}
H(n,\phi) = -\sum_{\langle x,y\rangle} n_x n_y \bigl[\cos (\phi _x-\phi _y) - 1\bigr]- \lambda \sum_{x}n_x-
\kappa\sum_{\langle x,y\rangle}n_x n_y.
\end{equation}
Here, as before, $\sigma_x\in\{1,\dots,q\}$ are the Potts spins,  $\phi _x\in [-\pi,\pi)$ are variables 
representing the ``angle'' of the corresponding $O(2)$-spins,
and $n_x\in \{0,1\} $ indicates the presence or absence of a particle (that carries the Potts spin $\sigma_x$ 
or the angle variable $\phi_x$) at site~$x$.

On the basis of ``usual'' arguments, the high temperature region is characterized by disordered configurations while 
the low temperatures features configurations with a strong (local) order, at least at small-to-intermediate dilutions.
The phenomenon discovered in \cite{CKS1,CKS2} is the existence of a region of intermediate temperatures and chemical
potentials, sandwiched  between the low temperature/high density ordered region
and the high temperature/low density disordered region, where typical configurations exhibit preferential occupation 
of one of the even/odd sublattices.
The appearance of such states is due to an \emph{effective entropic repulsion}.
Indeed, at low temperatures the spins on particles at neighboring sites are forced to be (nearly) aligned while if a particle is completely isolated, 
its spin is permitted to enjoy the full freedom of the available spin space.
Hence, at intermediate temperatures and moderate dilutions, 
there is an entropic advantage for the particles to occupy only one of the sublattices.

Let us concentrate on the portion of the phase boundary between the staggered region and the low temperature region.  
The claim can be stated uniformly for both systems in \twoeqref{E:Potts-dil-Ham}{E:XY-dil} provided we introduce 
the relevant good events in terms of occupation variable~$n$.
Namely, we let:
\begin{equation}
\label{E:dilP-ord-stagg}
\begin{aligned}
\GG^{\text{dense}}&=\bigl\{(\sigma,n)\colon n_x=1 \text{ for all } x\in\Lambda_1\bigr\},
\\
\GG^{\text{even}}&=\bigl\{(\sigma,n)\colon n_x=\1_{\{x\text{ even}\}} \text{ for all } x\in\Lambda_1\bigr\}, 
\\
\GG^{\text{odd}}&=\bigl\{(\sigma,n)\colon n_x=\1_{\{x\text{ odd}\}}  \text{ for all }x\in\Lambda_1\bigr\}.
\\\end{aligned}
\end{equation}
Again, using slightly modified versions of these events, it was shown in \cite{CKS1,CKS2} that 
there exist positive numbers $\epsilon,\kappa_0\ll1$ and, for every $\kappa \in(0,\kappa_0)$, an interval $I(\kappa)\subset \mathbb R$
such that the following is true: 
For any $\lambda\in I$ there exist inverse temperatures
$\beta_1(\kappa, \lambda)$ and $\beta_2(\kappa, \lambda)$, and a transition temperature 
$\betat(\kappa, \lambda)\in[\beta_1,\beta_2]$ such that
\begin{enumerate}
\item [(1)]
for any $\beta \in [\betat,\beta_2]$ there exists an ``densely occupied''  state
$\mu^{\text{dense}} \in \frakG_{\beta}$,
for which
\begin{equation}
\label{E:dilP-ord}
\rho_{\mu^{\text{dense}}}(\GG^{\text{dense}}) \ge 1-\epsilon,
\end{equation}
\item[(2)]
for any $\beta \in [\beta_1,\betat]$ there exist two states
$\mu^{\text{even}}, \mu^{\text{odd}}\in   \frakG_{\beta}$ satisfying
\begin{equation}
%\label{E:}
\rho_{\mu^{\text{even}}}(\GG^{\text{even}})
\ge 1-\epsilon   \quad \text{ and  }\quad  \rho_{\mu^{\text{odd}}}(\GG^{\text{odd}})
\ge 1-\epsilon.
\end{equation}
\end{enumerate} 
The error~$\epsilon$ is of order $\beta^{-\ffrac18}$ 
(cf.~the bound (2.15) in \cite{CKS2}) in the case of the $XY$-model in $d=2$, 
and it tends zero as $q\to\infty$ in the case of the diluted Potts model.

A somewhat stronger conclusion can be made for the diluted Potts model.
Namely, at~$\beta=\betat$, there are actually $q+2$ distinct states, two staggered states $\mu^{\text{even}}$ 
and $\mu^{\text{odd}}$ and~$q$ ordered states $\mu^{\text{dense},m}$, with the latter characterized 
by the condition
\begin{equation}
%\label{E:}
\rho_{\mu^{\text{dense},m}}(\GG^{\text{dense},m})
\ge 1-\epsilon,
\end{equation}
where 
\begin{equation}
%\label{E:}
\GG^{\text{dense}, m}=\bigl\{(\sigma, n)\colon n_x=1 \text{ and } \sigma_x=m \text{ for all } x\in
\Lambda_1\bigr\}.
\end{equation}
It is plausible that an analogous conclusion applies to the XY-model in~$d\ge3$ 
because there the low-temperature phase should exhibit magnetic order. 
However, in~$d=2$ such long-range order is not permitted by the Mermin-Wagner theorem 
and so there one expects to have only 3 distinct ergodic Gibbs states at~$\betat$.

A weaker form of the expected conclusion is an easy consequence of our Theorem~\ref{T:complete}: 
For each extremal 2-periodic Gibbs state
$\mu\in\frakG_{\betat}$  there exists $\GG\in\{\GG^{\text{even}},\GG^{\text{odd}},\GG^{\text{dense}}\}$
(in the case of diluted Potts model,  $\GG\in\{\GG^{\text{even}},\GG^{\text{odd}},\GG^{\text{dense},m}, m=1,\dots,q\}$)
such that
\begin{equation}
%\label{E:}
\rho_\mu(\GG)\ge 1-\epsilon.
\end{equation}
In particular, no ergodic Gibbs state $\mu\in\frakG_{\betat}$ has particle density in 
$[\epsilon,\ffrac12-\epsilon]\,\cup\,[\ffrac12+\epsilon,1-\epsilon]$. 
The proof of these observations goes by noting that the smallness of~$\pp_\beta(\BB)$ 
for the bad event  $\BB=(\GG^{\text{dense}}\cup\GG^{\text{even}}\cup\GG^{\text{odd}})^\cc$
is a direct consequence of the corresponding bounds from \cite{CKS1, CKS2} of the ``contour events.''
In the case of the XY-model in dimension $d=2$, this amounts to the bounds (2.9) and (2.15) from \cite{CKS2}.

\begin{remark}
A more general class of models, with spin taking values in a Riemannian manifold, is also considered in  \cite{CKS2}.
A related phase transition in an annealed diluted $O(n)$ Heisenberg ferromagnet has been proved in~\cite{CSZ}.
\end{remark}

\subsection{Order-by-disorder transitions}
Another class of systems where our results provide new information are the $O(2)$-nearest 
and next-nearest neighbor antiferromagnet~\cite{BCKiv}, the 120-degree model~\cite{BCN1}, 
and the orbital-compass model~\cite{BCN2}. All of these are continuum-spin systems whose common feature is that 
the infinite degeneracy of the ground states is broken, at positive temperatures, by long-wavelength (spin-wave) excitation.
We will restrict our attention to the first of these models, the~$O(2)$-nearest and next-nearest neighbor antiferromagnet.
The other two models are somewhat more complicated---particularly, 
due to the presence of non-translation invariant ground states---but the conclusions are fairly analogous.

Consider a spin system on~$\Z^2$ whose spins, $\bS_x$, take values on the unit circle in~$\R^2$ with \emph{a priori} uniform distribution.
The Hamiltonian is
\begin{equation}
\label{Ham}
H(\bS)=\sum_{x}\bigl(\bS_{x}\cdot\bS_{x+\hate_1+\hate_2}
+\bS_{x}\cdot\bS_{x+\hate_1-\hate_2}\bigr)
+\gamma\sum_{x}\bigl(\bS_{x}\cdot\bS_{x+\hate_1}
+\bS_{x}\cdot\bS_{x+\hate_2}\bigr),
\end{equation}
where~$\hate_1$ and~$\hate_2$ are the unit vectors in the coordinate lattice directions and the dot denotes the usual scalar product.
Note that both nearest and next-nearest neighbors are coupled antiferromagnetically but with a different strength.
The following are the ground state configurations for~$\gamma\in(-2,2)$: 
Both even and odd sublattices enjoy a Ne\'el (antiferromagnetic) order, but the relative orientation of these sublattice states is arbitrary.

It is clear that, at low temperatures, the configurations will be locally near one of the aforementioned ground states.
Due to the continuous nature of the spins, the fluctuation spectrum is dominated by ``harmonic perturbations,''
a.k.a.~\emph{spin waves}.
A heuristic spin-wave calculation (cf.~\cite[Sect.~2.2]{BCN1} for an example in the context of the 120-degree model) suggests 
that among all~$2\pi$ possible relative orientations of the sublattices, 
the parallel and the antiparallel orientations are those entropically most favorable.
And, indeed, as was proved in~\cite{BCKiv}, there exist two $2$-periodic Gibbs states~$\mu_1$ and~$\mu_2$ 
with the corresponding type of long-range order.
However, the existence of Gibbs states with other relative orientations has not been ruled out.

We will now state a stronger version of~\cite[Theorem~2.1]{BCKiv}.
Let~$B$ be a large even integer and consider two~$B$-block events~$\GG_1$ and~$\GG_2$ defined as follows:
fixing a positive~$\kappa\ll1$,  let
\begin{equation}
%\label{}
\GG_1=\bigcap_{\begin{subarray}{c}
x,y\in\Lambda_B\\(y-x)\cdot\hate_2=0
\end{subarray}}
\{\bS_x\cdot\bS_y\ge1-\kappa\}
\cap
\bigcap_{x,x+\hate_2\in\Lambda_B}
\{\bS_x\cdot\bS_{x+\hate_2}\le-1+\kappa\},
\end{equation}
i.e., $\GG_1$ enforces horizontal stripes all over~$\Lambda_B$.
The event~$\GG_2$ in turn enforces vertical stripes; the definition is as above with the roles of~$\hate_1$ and~$\hate_2$ interchanged. 
Then we have:

\begin{theorem}
\label{T:AF}
Let~$\gamma\in(0,2)$ and let~$\kappa\ll1$. For each~$\epsilon>0$ there exists  $\beta_0\in(0,\infty)$ such that 
for each~$\beta\ge\beta_0$:
\begin{enumerate}
\item[(1)]
There exist two ergodic Gibbs states~$\mu_1,\mu_2\in\frakG_\beta$, such that
\begin{equation}
\label{4.11d}
\rho_{\mu_j}(\GG_j)\ge1-\epsilon,\qquad j=1,2.
\end{equation}
\item[(2)] There exists an integer~$B\ge1$ such that for any~$\mu\in\frakG_\beta$ 
that is ergodic with respect to shifts by multiples of~$B$ we have
\begin{equation}
%\label{}
\text{ either } \ \rho_\mu(\GG_1)\ge 1-\epsilon  \ \text{ or }\    \rho_\mu(\GG_2)\ge 1-\epsilon.
\end{equation}
\end{enumerate}
\end{theorem}

The first conclusion---the existence of Gibbs states with parallel and antiparallel relative orientation 
of the sublattices---was the main content of Theorem~2.1 of~\cite{BCKiv}.
What we have added here is that the corresponding configurations dominate \emph{all} ergodic Gibbs states.
The~$O(2)$ ground-state symmetry of the relative orientation of the sublattices is thus truly broken at positive temperatures, 
which bolsters significantly the main point of~\cite{BCKiv}.
Note that no restrictions are posed on the overall orientation of the spins. 
Indeed, by the Mermin-Wagner theorem every $\mu\in\frakG_\beta$ is invariant 
under simultaneous rotations of all spins. 

\begin{proofsect}{Proof of Theorem~\ref{T:AF}}
As expected, the proof boils down to showing that, for a proper choice of scale~$B$ we have $\pp_\beta(\BB)\ll1$ 
for~$\BB=(\GG_1\cup\GG_2)^\cc$.
In~\cite{BCKiv} this is done by decomposing~$\BB$ into more elementary events---depending on whether the ``badness'' 
comes from excessive energy or insufficient entropy---and estimating each of them separately.
The relevant bounds are proved in \cite[Lemmas~4.4 and~4.5]{BCKiv} and combined together in~\cite[Eq.~(4.20)]{BCKiv}.
Applying Theorem~\ref{T:complete} of the present paper, we thus know that every $B$-shift ergodic~$\mu\in\frakG_\beta$ 
is dominated either by blocks of type~$\GG_1$ or by blocks of type~$\GG_2$. Since~$\rho_\mu(\BB)\le\epsilon$ in all states, 
the existence of~$\mu_1,\mu_2\in\frakG_\beta$ satisfying \eqref{4.11d} 
follows by symmetry with respect to rotation (of the lattice) by~$90$-degrees.
\end{proofsect}

\subsection{Nonlinear vector models}
A class of models with continuous symmetry that are conceptually close to the Potts model
has been studied recently by van~Enter and Shlosman~\cite{ES}.
As for our previous examples with continuous spins,  Pirogov-Sinai theory is not readily available
and one has to rely on chessboard estimates. We will focus our attention
on one example in this class, a \emph{nonlinear
ferromagnet}, although our conclusions apply with appropriate, and somewhat delicate, modifications
also to liquid crystal models and lattice gauge models
discussed in~\cite{ES}.  

Let us consider an $O(2)$-spin system on $\Z^2$ 
with spins parametrized by the angular variables $\phi_x\in(-\pi,\pi]$.
The Hamiltonian is given by
\begin{equation}
\label{E:VF-Ham}
H(\phi)=-\sum_{\langle x,y\rangle}\Bigl(\frac{1+\cos(\phi_x-\phi_y)}{2}\Bigr)^p,
\end{equation}
where $p$ is a nonlinearity parameter.
The \emph{a priori} distribution of the $\phi_x$'s is the Lebesgue measure on $(-\pi,\pi]$; 
the difference $\phi_x-\phi_y$ is always taken modulo~$2\pi$.

In order to define the good block events, we first split all bonds into three classes. Namely, given a configuration 
$(\phi_x)_{x\in \Z^2}$, we say that the bond $\langle x,y\rangle$ is 
\begin{enumerate}
\item[(1)] \emph{strongly ordered} if $|\phi_x-\phi_y|\le \frac1{C\sqrt p}$,
\item[(2)] \emph{weakly ordered} if $\frac1{C\sqrt p}<|\phi_x-\phi_y|< \frac{C}{\sqrt p}$, and
\item[(3)] \emph{disordered} if $|\phi_x-\phi_y|\ge  \frac{C}{\sqrt p}$.
\end{enumerate}
Here $C$ is a large number to be determined later. 
If a bond is either strongly or weakly ordered, we will call it simply \emph{ordered}.

On the basis of \eqref{E:VF-Ham}, it is clear that strongly ordered bonds are favored energetically 
while the disordered bonds are favored entropically. The main observation of \cite{ES}---going back to~\cite{DS,Kotecky-Shlosman,Alexander-Chayes}---is that,
at least in torus measures, ordered and disordered bonds are unlikely to occur in the same configuration.
This immediately implies coexistence of at least two distinct states at some intermediate temperature. Moreover, since it is also unlikely to have many bonds in the ``borderline'' region $|\phi_x-\phi_y|\approx \frac{C}{\sqrt p}$, the transition is accompanied by a jump in the energy density. But, to prove that the energy gap stays uniformly positive as~$p\to\infty$, it appears that one needs to establish the existence of a free-energy barrier between the
\emph{strongly} ordered and disordered phases.

\newcommand{\so}{{\text{\rm so}}}
\newcommand{\wo}{{\text{\rm wo}}}
\newcommand{\mix}{{\text{\rm mix}}}
\newcommand{\dis}{{\text{\rm dis}}}

Let~$\Lambda_1$ be a $1$-block (i.e., a plaquette) and let us consider the following good events on~$\Lambda_1$:
The event that all bonds on~$\Lambda_1$ are strongly ordered,
\begin{equation}
%\label{}
\GG_\so=\Bigl\{|\phi_x-\phi_y|\le \frac1{C\sqrt p}\colon\,\forall x,y\in\Lambda_1,\,|x-y|=1\Bigr\},
\end{equation}
and the event that all bonds on~$\Lambda_1$ are disordered,
\begin{equation}
%\label{}
\GG_\dis=\Bigl\{|\phi_x-\phi_y|\ge \frac C{\sqrt p}\colon\,\forall x,y\in\Lambda_1,\,|x-y|=1\Bigr\}.
\end{equation}
Then we have:

\begin{theorem}
\label{thm4.1}
For each~$\epsilon>0$ and each sufficiently large~$C>1$, there exists~$p_0>0$ such that for all~$p>p_0$, 
there exists a number~$\betat\in(0,\infty)$ and two distinct, 
shift-ergodic Gibbs states~$\mu^\so,\mu^\dis\in\frakG_{\betat}$ such that
\begin{equation}
\label{4.16c}
\rho_{\mu^\so}(\GG_\so)\ge1-\epsilon
\quad\text{and}\quad
\rho_{\mu^\dis}(\GG_\dis)\ge1-\epsilon.
\end{equation}
In addition, for all shift-ergodic Gibbs states~$\mu\in\frakG_{\beta_t}$, we have
\begin{equation}
\label{4.14c}
\text{ either }\ \rho_\mu(\GG_\dis)\ge 1-\epsilon \ \text{ or }\ 
\rho_\mu(\GG_\so) \ge 1-\epsilon,
\end{equation}
while
\begin{equation}
\label{4.14aa}
\rho_\mu(\GG_\so) \ge 1-\epsilon\ \text{ for all shift-ergodic }\ \mu\in\frakG_{\beta}\ \text{ with }\ \beta>\beta_t
\end{equation}
 and 
\begin{equation}
\label{4.14bb}
\rho_\mu(\GG_\dis) \ge 1-\epsilon \ \text{ for all shift-ergodic }\ \mu\in\frakG_{\beta} \ \text{ with }\ \beta<\beta_t.
\end{equation}
Finally, for every~$p>p_0$ and~$C$ large, every ergodic Gibbs state will have energy near zero when~$\beta>\betat$ 
and at least~$1-O(C^{-2})$ when~$\beta<\betat$.
\end{theorem}

We remark that the existence of a first-order transition in energy density has been a matter of some controversy in the physics literature; 
see~\cite{ES-PRL,ES} for more discussion and relevant references.
The proof of Theorem~\ref{thm4.1} is fairly technical and it is therefore deferred to Sect.~\ref{sec5}.

\subsection{Magnetostriction transition}
Our final example is the magnetostriction transition studied recently by Shlosman and Zagrebnov~\cite{SZ}. 
The specific system considered in~\cite{SZ} has the Hamiltonian 
\begin{equation}
\label{E:SZ-Ham}
H(\sigma,r)=-\sum_{\langle  x,y \rangle}J(r_{x,y})\sigma_x\sigma_y +\kappa \sum_{\langle  x,y \rangle}(r_{x,y}-R)^2
+\lambda \!\!\!\sum_{\begin{subarray}{c} \langle  x,y \rangle, \langle  z,y \rangle\\
 |x-z|=\sqrt 2\end{subarray}}\!\!\!(r_{x,y}-r_{z,y})^2.
\end{equation}
Here the sites $x\in \Z^d$ label the atoms in a crystal; the atoms have magnetic moments represented by the Ising spins~$\sigma_x$. 
The crystal is not rigid; the variables $r_{x,y}\in\R$, $r_{x,y}>0$, play the role of spatial distance between neighboring   crystal sites.

The word \emph{magnetostriction} refers to the phenomenon where a solid undergoes a magnetic transition 
accompanied by a drastic change in the crystalline structure.
In~\cite{SZ} such a transition was proven for interaction potentials $J=J(r_{x,y})$ that are strong at short distances and weak at large distances.
The relevant states are characterized by disjoint \emph{contracted},  
\begin{equation}
%\label{}
\GG^{\text{contr}}=\bigl\{(r,\sigma)\colon r_{x,y}\le \eta,\,\forall x, y\in\Lambda_1, |x-y|=1\bigr\},
\end{equation}
and \emph{expanded},
\begin{equation}
%\label{}
\GG^{\text{exp},\pm}=\bigl\{(r,\sigma)\colon r_{x,y}\ge \eta+\epsilon,\,\forall x, y\in\Lambda_1, |x-y|=1\bigr\}
\cap\bigl\{\sigma_x=\pm1,\,\forall x\in\Lambda_1\bigr\},
\end{equation}
block events.
The parameters $\eta$ and $\varepsilon$ can be chosen so that there exists~$\betat\in(0,\infty)$ for which the following holds:
\begin{enumerate}
\item[(1)] 
For all $\beta\le \betat$ there exists an \emph{expanded} Gibbs state $\mu^{\text{exp}}\in\frakG_\beta$ such that
$\rho_{\mu^{\text{exp}}}(\GG^{\text{exp}})\ge \ffrac34$;
\item[(2)] 
For all $\beta\ge \betat$ there exist two distinct \emph{contracted} Gibbs states 
$\mu^{\text{contr},\pm}\in\frakG_\beta$ such that 
$\rho_{\mu^{\text{contr},\pm}}(\GG^{\text{contr},\pm})\ge \ffrac34$.
\end{enumerate}
In particular at $\beta=\betat$ there exist three distinct Gibbs states; one expanded and two contracted 
with opposite values of the magnetization. The authors conjecture that  these are the only shift-ergodic Gibbs states at $\beta=\betat$.

Unfortunately, the above system has unbounded interactions and so it is not strictly 
of the form for which Theorem~\ref{T:complete} applies. Instead we will use the following generalization:

\begin{theorem}
\label{thm4.2}
Let~$d\ge2$ and consider a spin system with translation-invariant finite-range 
interaction potentials $(\Phi_A)_{A\Subset\Z^d}$ such 
that the torus measure is reflection positive for all even~$L$. Let~$\GG_1,\dots,\GG_r$ be a collection of good~$B$-block events 
satisfying the requirements in Theorem~\ref{T:complete} and let~$\BB$ be the corresponding bad event. 
Then for all~$\epsilon>0$ there exists~$\delta>0$---depending possibly only on~$d$ but not on details of the model nor on~$n$ or~$B$---such that for all~$\beta\ge0$ for which~$\pp_\beta(\BB)<\delta$ the following is true: 
If~$\mu\in\frakG_\beta$ is a $B$-shift ergodic Gibbs state with
\begin{equation}
\label{E:ener}
\sum_{\begin{subarray}{c}
A\colon A\Subset\Z^d\\0\in A
\end{subarray}}
E_\mu\bigl(|\Phi_A|\bigr)<\infty,
\end{equation}
then we have
\begin{equation}
\label{E:alt1a}
\rho_\mu(\BB)\in [0,\epsilon]
\end{equation}
and there exists~$i\in\{ 1, \dots, r \}$
 such that
\begin{equation}
\label{E:alt2a}
\rho_\mu(\GG_i)\ge 1-\epsilon.
\end{equation}
\end{theorem}
 
\begin{proofsect}{Proof}
The proof is virtually identical to that of Theorem~\ref{T:complete} with one exception: Since the interactions are not bounded, 
we cannot use Lemma~\ref{lemma3.4} directly. Suppose we have a Gibbs state~$\mu$ 
that obeys \eqref{E:ener} 
but violates one of the conditions \twoeqref{E:alt1a}{E:alt2a}. Let~$R_{M,N}(\CC_N)$ be as in \eqref{E:R_MN}. 
Lemma~\ref{lemma3.1} still applies and so we have \eqref{E:R_MN>} for some~$N$. 

Let $L=MNB$ and let~$\DD_M$ be the event that the boundary energy in the box~$\Lambda$ is less than~$cM^{d-1}$, i.e.,
\begin{equation}
%\label{}
\DD_M=\biggl\{\sum_{\begin{subarray}{c}
A\colon A\cap\Lambda_L\ne\emptyset\\A\cap\Lambda_L^\cc\ne\emptyset
\end{subarray}}
|\Phi_A|\le cM^{d-1}\biggr\}.
\end{equation}
where~$c$ is a positive constant. In light of the condition \eqref{E:ener}, the fact that the interaction has a finite range, 
and the Chebyshev bound, it is clear that we can choose~$c$ so that $\mu(\DD_M^\cc)<(4N^d)^{-1}$ for all~$M$. Hence, we have
\begin{equation}
%\label{}
\mu\bigl(\DD_M\cap\{R_{M,N}(\CC_N)>\ffrac12\}\bigr)>\frac1{4N^d}.
\end{equation}
Next let~$s$ and~$s'$ be as in the proof of Lemma~\ref{lemma3.4} and suppose that both~$s$ and~$s'$ belong to~$\DD_M$. 
Then, by definition,
\begin{equation}
%\label{}
\bigl|H_{\Lambda_L}(s)-H_{\Lambda_L}(s')\bigr|\le 2cM^{d-1}
\end{equation}
and, applying the rest of the proof of Lemma~\ref{lemma3.4}, we thus have
\begin{equation}
%\label{}
\mu\bigl(\DD_M\cap\{R_{M,N}(\CC_N)>\ffrac12\}\bigr)
\le
\texte^{2\beta cM^{d-1}}
\BbbP_{2L,\beta}\bigl(\DD_M\cap\{R_{M,N}(\CC_N)>\ffrac12\}\bigr).
\end{equation}
Neglecting $\DD_L$ on the right-hand side and invoking \eqref{3.29}, we again derive the desired contradiction once~$M$ is sufficiently large.
\end{proofsect}

With Theorem~\ref{thm4.2} in the hand, we can extract the desired conclusion for the magnetostriction transition.
First, the energy condition is clearly satisfied in any state generated by tempered boundary conditions. 
We then know that, in every such ergodic state~$\mu$, only a small number blocks will feature bonds that are 
neither contracted (and magnetized) nor expanded (and non-magnetized):
\begin{equation}
\label{E:SZ-rho}
\rho_\mu(\GG^{\text{exp}}),  \rho_\mu(\GG^{\text{exp},\pm}) \in [0,\epsilon]\,\cup\,[1-\epsilon,1] \quad
\text{ and } \quad\rho_\mu(\BB)\le\epsilon.
\end{equation}

The existence of a phase transition follows by noting that the contracted states have less energy than the expanded ones; 
there is thus a jump in the energy density as the temperature varies.

%\newpage
\section{Appendix}
\label{sec5}\noindent
The goal of this section is to prove Theorem~\ref{thm4.1} which concerns the non-linear vector model with interaction \eqref{E:VF-Ham}. 
The technical part of the proof is encapsulated into the following claim:

\begin{proposition}
\label{prop5.1}
There exists a constant~$C_0>0$ such that for all~$\delta>0$ and all~$C\ge C_0$ the following holds: There exists~$p_0>0$ such 
that for all~$p\ge p_0$ we have
\begin{equation}
\label{5.1}
\sup_{\beta\ge0}\,\pp_\beta((\GG_\so\cup\GG_\dis)^\cc)<\delta
\end{equation}
and
\begin{equation}
\label{5.2}
\lim_{\beta\to\infty}\pp_\beta(\GG_\dis)=0
\quad\text{and}\quad
\lim_{\beta\downarrow0}\,
\pp_\beta(\GG_\so)<\delta.
\end{equation}
\end{proposition}

To prove this proposition, we will need to carry out a sequence of energy and entropy bounds.
To make our energy estimates easier, and uniform in~$p$, we first notice that there are constants~$0<a<b$ such that
\begin{equation}
\label{E:enerbd}
\texte^{-bx^2}\le\frac{1+\cos(x)}2\le\texte^{-ax^2},
\qquad -1\le x\le1.
\end{equation}
The argument commences by splitting the bad event~$\BB=(\GG_\so\cup\GG_\dis)^\cc$ into two events: 
The event~$\BB_\wo$ that $\Lambda_1$ 
contains a weakly-ordered bond, and~$\BB_\mix=\BB\setminus\BB_\wo$ which, as a moment's thought reveals, 
is the event that~$\Lambda_1$ contains two adjacent bonds one of which is strongly ordered and the other disordered.
The principal chessboard estimate  yields the following lemma:

\begin{lemma}
\label{lemma5.2}
Suppose that $C\le\sqrt p$. Then
\begin{equation}
\label{4.15}
\pp_\beta(\BB_\wo)\le4\,\biggl(\min\Bigl\{\tfrac{C^2}\kappa\,\texte^{-2\beta[\texte^{-{b\kappa^2}/{C^2}}-
\texte^{- a/{C^2}}]},\,\tfrac C{\pi\sqrt p}\,\texte^{2\beta\texte^{-a/{C^2}}}\Bigr\}\biggr)^{\ffrac14}
\end{equation}
and
\begin{equation}
\label{4.16}
\pp_\beta(\BB_\mix)\le4\biggl(\min\Bigl\{\texte^{-2\beta[\frac32\texte^{-{b}/{C^2}}-1-\texte^{-aC^2}]},\,
\texte^{2\beta}\bigl(\tfrac1{\pi C\sqrt p}\bigr)^{\ffrac34}
\Bigr\}\biggr)^{\ffrac12}
\end{equation}
 for all~$\beta\ge0$ and all~$\kappa\in(0,1)$.
Moreover, we have
\begin{equation}
\label{4.15a}
\pp_\beta(\GG_\dis)\le
\pi C\sqrt p\,\exp\bigl\{-2\beta[\texte^{-\frac b{C^2}}-\texte^{-aC^2}]\bigr\}
\end{equation}
and
\begin{equation}
\label{4.16a}
\pp_\beta(\GG_\so)\le\frac1\pi\frac{\texte^{2\beta}}{C\sqrt p}.
\end{equation}
\end{lemma}

\begin{proofsect}{Proof}
Let~$Z_L$ be the partition function obtained by integrating~$\texte^{-\beta H_L}$ over all allowed configurations.
Consider the following reduced partition functions: 
%\settowidth{\leftmargini}{(11)}
\begin{enumerate}
\item[(1)]
$Z_L^\dis$, obtained by integrating~$\texte^{-\beta H_L}$ subject to the restriction that every bond in~$\T_L$ is disordered.
\item[(2)]
$Z_L^\so$, obtained similarly while stipulating that every bond in~$\T_L$ is strongly ordered.
\item[(3)]
$Z_L^\wo$, in which every bond in~$\T_L$ is asked to be weakly ordered.
\item[(4)]
$Z_L^\mix$, enforcing that every other horizontal line contains only strongly-ordered bonds, and the remaining lines contain 
only disordered bonds. A similar periodic pattern is imposed on vertical lines as well.
\end{enumerate}
To prove the lemma, we will need upper and lower bounds on the partition functions in (1-2), 
and upper bounds on the partition functions in~(3-4).

We begin by upper and lower bounds on~$Z_L^\dis$.
First, using the fact that the Hamiltonian is always non-positive, we have~$\texte^{-\beta H_L}\ge1$.
On the other hand, the inequalities \eqref{E:enerbd} and a natural monotonicity of the interaction imply that 
\begin{equation}
%\label{}
\Bigl(\frac{1+\cos(\phi_x-\phi_y)}{2}\Bigr)^p
\le \Bigl(\frac{1+\cos(C/\sqrt p)}{2}\Bigr)^p
\le \texte^{-a C^2}
\end{equation}
whenever~$\langle x,y\rangle$ is a disordered bond.
In particular,~$-\beta H_L$ is less than~$2\beta\texte^{-a C^2}|\T_L|$ for every configuration contributing to~$Z_L^\dis$.
Using these observations we now easily  derive that
\begin{equation}
\label{4.19}
(2\pi)^{|\T_L|}\le Z_L^\dis\le (2\pi)^{|\T_L|}\,\texte^{2\beta\texte^{-aC^2}|\T_L|}.
\end{equation}
Similarly, for the partition function~$Z_L^\so$ we get
\begin{equation}
\label{4.20}
\Bigl(\texte^{2\beta\texte^{-{b\kappa^2}/{C^2}}}\frac {2\kappa}{C\sqrt p}\Bigr)^{|\T_L|}
\le Z_L^\so\le 2\pi \texte^{2\beta|\T_L|}
\Bigl(\frac2{C\sqrt p}\Bigr)^{|\T_L|-1}.
\end{equation}
Indeed, for the upper bound we first note that~$-\beta H_L\le2\beta|\T_L|$. 
Then we fix a tree spanning all vertices of~$\T_L$, disregard the constraints everywhere 
except on the edges in the tree and, starting from the ``leaves,'' we sequentially integrate all site variables. 
(Thus, each site is effectively forced into an interval of length~$\frac2{C\sqrt p}$, 
except for the ``root'' which retains all of its~$2\pi$ possibilities.) 
For the lower bound we fix a number~$\kappa\in(0,1)$ and restrict the integrals to configurations such that 
$|\phi_x-\phi_y|\le\frac\kappa{C\sqrt p}$ for all bonds~$\langle x,y\rangle$ in~$\T_L$. 
The bound~$-\beta H_L\ge2\beta\texte^{-{b\kappa^2}/{C^2}}|\T_L|$ then permits us to estimate away 
the Boltzmann factor for all configurations; the entropy factor reflects the fact that each site can vary throughout an interval 
of length at least~$\frac{2\kappa}{C\sqrt p}$.

Next we will derive good upper bounds on the remaining two partition functions. 
First, similar estimates as those leading to the upper bound in \eqref{4.20} give us
\begin{equation}
\label{4.17}
Z_L^\wo\le2\pi\Bigl(\texte^{2\beta\texte^{-a/C^2}}\frac{2C}{\sqrt p}\Bigr)^{|\T_L|}.
\end{equation}
For the partition function~$Z_L^\mix$ we note that $\ffrac14$ of all sites are adjacent only to disordered bonds, 
while the remaining~$\ffrac34$ are connected to one another via a grid of strongly-ordered bonds.
Estimating~$-\beta H_L\le\beta(1+\texte^{-aC^2})|\T_L|$ for all relevant configurations, 
similar calculations as those leading to~\eqref{4.20} again give us
\begin{equation}
\label{4.18}
Z_L^\mix\le 2\pi
\texte^{\beta(1+\texte^{-aC^2})|\T_L|}\,
(2\pi)^{\frac{|\T_L|}4}\,
\Bigl(\frac2{C\sqrt p}\Bigr)^{\frac34|\T_L|-1}.
\end{equation}
It now remains to combine these estimates into the bounds on the quantities on the left-hand side of \twoeqref{4.15}{4.16} 
and \twoeqref{4.15a}{4.16a}.

We begin with the bound \eqref{4.15a}. Clearly,~$\pp_\beta(\GG_\dis)$ is the $L\to\infty$ limit of~$(Z_L^\dis/Z_L)^{1/{|\T_L|}}$, 
which using the lower bound~$Z_L\ge Z_L^\so$ with~$\kappa=1$ easily implies \eqref{4.15a}.
The bound \eqref{4.16a} is obtained similarly, except that now we use that~$Z_L\ge Z_L^\dis$. 
The remaining two bounds will conveniently use the fact that for two-dimensional nearest-neighbor models, and square tori, 
the torus measure~$\BbbP_{L,\beta}$ is reflection positive even with respect to the diagonal planes in~$\T_L$.
Indeed, focusing on \eqref{4.15} for a moment, we first note that~$\BB_\wo$ is covered by the union of four (non-disjoint) events 
characterized by the position of the weakly-ordered bond on~$\Lambda_1$. If~$\BB_\wo^{(1)}$ is the event 
that the lower horizontal bond is the culprit, 
the subadditivity property of~$\pp_\beta$---see Lemma~6.3 of~\cite{BCN1}---gives us~$\pp_\beta(\BB_\wo)\le4\pp_\beta(\BB_\wo^{(1)})$.
 Disseminating~$\BB_\wo^{(1)}$ using reflections in coordinate directions, we obtain an event enforcing weakly-ordered bonds 
on every other horizontal line. 
Next we apply a reflection in a diagonal line of even parity to make this into an even parity grid.
From the perspective of reflections in odd-parity diagonal lines---i.e., 
those not passing through the vertices of the grid---half of the ``cells'' enforces all four bonds therein to be weakly ordered, 
while the other half does nothing.
Applying chessboard estimates for these diagonal reflections, we get rid of the latter cells.
The result of all these operations is the bound
\begin{equation}
%\label{}
\pp_\beta(\BB_\wo)\le\lim_{L\to\infty}4\Bigl(\frac{Z_L^\wo}{Z_L}\Bigr)^{\frac1{4|\T_L|}}.
\end{equation}
Estimating~$Z_L$ from below by the left-hand sides of \twoeqref{4.19}{4.20} now directly implies \eqref{4.15}.

The event~$\BB_\mix$ is handled similarly: First we fix a position of the ordered-disordered pair of bonds and use subadditivity 
of~$\pp_\beta$ to enforce the \emph{same} choice at every lattice plaquette; this leaves us with four overall choices.
 Next we use diagonal reflections to produce the event underlying~$Z_L^\mix$.
Estimating~$Z_L$ from below by $\ffrac14$-th power of the lower bound in \eqref{4.19} and $\ffrac34$-th power of the lower bound
 in \eqref{4.20} with~$\kappa=1$, we get the first term in the minimum in~\eqref{4.16}.
To get the second term, we use that~$Z_L\ge Z_L^\dis$, apply \eqref{4.18} and invoke the bound~$1+\texte^{-aC^2}\le2$.
\end{proofsect}

\begin{proofsect}{Proof of Proposition~\ref{prop5.1}}
The desired properties are simple consequences of the bounds in Lemma~\ref{lemma5.2}.
Indeed, if~$C$ is so large that $\texte^{-b/C^2}>\texte^{-aC^2}$, then \eqref{4.15a} implies 
that~$\pp_\beta(\GG_\dis)\to0$ as~$\beta\to\infty$.
On the other hand, \eqref{4.16a} shows that the~$\beta\to0$ limit of~$\pp_\beta(\GG_\so)$ is order~$\ffrac1{\sqrt p}$, 
which can be made as small as desired by choosing~$p$ sufficiently large.

To prove also~\eqref{5.1}, we first invoke Lemma~6.3 of~\cite{BCN1} one last time to see 
that~$\pp_\beta(\BB)\le\pp_\beta(\BB_\wo)+\pp_\beta(\BB_\mix)$.
We thus have to show that both~$\pp_\beta(\BB_\wo)$ and~$\pp_\beta(\BB_\mix)$ can be made arbitrary small by increasing~$p$ 
appropriately.
We begin with~$\pp_\beta(\BB_\mix)$.
Let~$C$ be so large that
\begin{equation}
%\label{}
\tfrac32\texte^{-{b}/{C^2}}-1-\texte^{-aC^2}>0.
\end{equation}
Then for~$\beta$ such that~$\texte^{2\beta}>p^{\ffrac14}$ the first term in the minimum in \eqref{4.15a} decays like a negative power 
of~$p$, while for the complementary values of~$\beta$, the second term is~$O(p^{-\ffrac18})$.
As to the remaining term, $\pp_\beta(\BB_\wo)$, here we choose~$\kappa\in(0,1)$ such that
\begin{equation}
%\label{}
\texte^{-b\kappa^2/C^2}-\texte^{-a/C^2}>0,
\end{equation}
and apply the first part of the minimum in \eqref{4.15} for~$\beta$ with~$\texte^{2\beta}\ge\sqrt p$, 
and the second part for the complementary~$\beta$, to show that~$\pp_\beta(\BB_\wo)$ is also bounded by constants time a negative power 
of~$p$, independently of~$\beta$.
Choosing~$p$ large, \eqref{5.1} follows.
\end{proofsect}

Now we can finally prove Theorem~\ref{thm4.1}:

\begin{proofsect}{Proof of Theorem~\ref{thm4.1}}
We will plug the claims of Proposition~\ref{prop5.1} in our main theorem. 
First, it is easy to check that the good block events~$\GG_\so$ and~$\GG_\dis$ satisfy 
the conditions~(1) and~(2) of Theorem~\ref{T:complete}.
Then \eqref{5.1} and \twoeqref{E:alt1}{E:alt2} imply that 
\begin{equation}
\label{E:konec}
\text{ either }\ \rho_\mu(\GG_\dis)\ge 1-\epsilon\ \text{ or }\  \rho_\mu(\GG_\so) \ge 1-\epsilon
\end{equation}
for all shift-ergodic Gibbs states~$\mu\in\frakG_{\beta}$ and all $\beta\in(0,\infty)$.
The limits \eqref{5.2} and Corollary~\ref{C:coex} then imply the existence of the transition temperature~$\betat$ 
and of the corresponding coexisting states.
Since  the energy density with negative sign undergoes a jump at~$\betat$ from values~$\gtrapprox\texte^{-b/C^2}$ 
to values~$\lessapprox\texte^{-aC^2}$---which differ by almost one once~$C\gg1$---all ergodic states 
for~$\beta>\betat$ must have small energy density while the states for~$\beta<\betat$ will have quite a lot of energy. 
Applying \eqref{E:konec}, all ergodic~$\mu\in\frakG_\beta$ for~$\beta>\betat$ must be dominated by strongly-ordered bonds, 
while those for~$\beta<\betat$ must be dominated by disordered bonds.
\end{proofsect}

\section*{Acknowledgments}
\noindent 
The research of M.B.~was supported by the NSF grant~DMS-0306167
and that of R.K.~by the grants GA\v CR  201/03/0478  and  MSM~0021620845. 
Large parts of this paper were written  while both authors visited Microsoft Research in Redmond.
The authors would like to thank Senya Shlosman, Aernout van Enter and an anonymous referee for many valuable suggestions on the first version of this paper.

%\newpage

\end{document}